# Note:Bridging Information Science: AB Initio Calculation Vortex of 2D Materials of Bismuthene(Bismuth Molecule) Graphene-Shaped through Kohn-Sham Equations

Yasuko Kawahata [†]

Faculty of Sociology, Department of Media Sociology, Rikkyo University, 3-34-1 Nishi-Ikebukuro,Toshima-ku, Tokyo, 171-8501, JAPAN.

ykawahata@rikkyo.ac.jp

**Abstract:** This study delves into the intricate electronic and optical behaviors of two-dimensional (2D) honeycomb materials, such as Stannen, arsenene, antimonene, silicene, and bismuthene(bismuth molecule), through the lens of first-principles calculations(AB Initio Calculations) based on the Kohn-Sham equations. Focusing on the exchange-correlation potential approximations within the Density Functional Theory (DFT) framework, we evaluate the potential of these materials in digital information control and management. Special attention is given to the nonlinear optical responses and electronic properties under the influence of twisted bilayer configurations, external fields, and varying twist angles. The findings offer novel insights into the design of advanced digital devices, suggesting a transformative approach to information technology through the utilization of 2D honeycomb materials.

## 1. Introduction

First-principles calculations(AB Initio Calculation) in condensed matter physics and materials science, as well as simulations of large-scale computations and chemical reaction processes, are powerful techniques for understanding and predicting the behavior of materials at the atomic and molecular levels. These computational methods are based on the fundamental laws of materials and can predict material properties without relying on experimental data. The significance of applying such methods to the diffusion processes of information in digital space is considered to be very substantial.

## 2. Understanding and Predicting the Flow of Information in Digital Space

The flow of information in digital space, especially in social media and online communities, is a highly complex system where numerous individual elements (users, posts, messages, etc.) interact. By applying methods from first-principles calculations, these interactions can be modeled based on fundamental laws, and the diffusion patterns and behavior of information flow can be predicted.

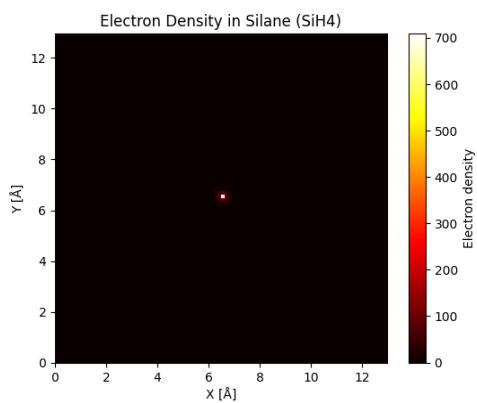

Fig. 1: Electron Density in Mono-Silane (SiH4)

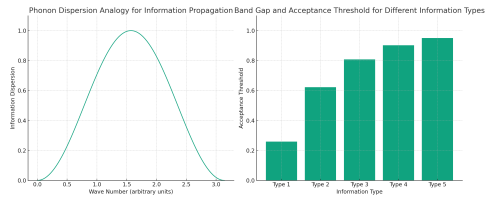

Fig. 2: Phonon Dispersion Analogy for Information Propagation, Band Gap and Acceptance Threshold for Different Information Types

## 3. Overcoming Undefined Patterns and Predictive Difficulty

Information flow in digital space is constantly changing, with new trends and streams of opinion emerging and fading away. These dynamics are often non-intuitive, making predictions challenging. By applying methods such as first-principles calculations in condensed matter physics and materials science, it might be possible to analyze these complex phenomena from first principles and reveal the emergence of new patterns and diffusion mechanisms.

## 4. New Paradigms in Understanding Digital Space

The application of first-principles calculations may provide a new theoretical framework for understanding the flow of information in digital space. Just as insights gained from analyzing the behavior of materials at the atomic level revolutionized materials science and condensed matter physics, reimagining the diffusion process of information as "digital particles" and understanding their interactions could offer new perspectives in information science and communication theory.

The application of first-principles calculations and large-scale computational methods from condensed matter physics and materials science to the diffusion processes of information in digital space holds great significance. This may enable us to understand the fundamental principles of complex and dynamic information flow and make predictions of new patterns and strategies for controlling information flow. It could be a step towards deeper understanding of digital space and building a better information environment.

In the rapidly evolving field of information technology, the exploration of novel materials with unique electronic and optical properties has become increasingly crucial for the development of next-generation digital devices and systems. Among these materials, two-dimensional (2D) honeycomb structures, such as antimonene, tellurene, silane, monosilane, silicene, and bismuthene, have garnered significant attention due to their remarkable characteristics that stem from their quantum mechanical nature. This paper delves into the first-principles calculations based on the Kohn-Sham equations within the framework of density functional theory (DFT) to elucidate the electronic structure and exchange-correlation potential approximations of these materials.

## 5. Overcoming Undefined Patterns and Predictive Difficulty

The flow of information in digital space forms highly complex dynamic systems due to its nonlinearity and the interaction of diverse elements. In these systems, small changes can lead to significant outcomes, and phenomena that cannot be captured by conventional predictive models may arise. In such situations, applying methods from condensed matter physics and materials science, such as first-principles calculations, can reveal the fundamental laws governing how individual units of information interact with each other. This involves constructing propagation models of information and introducing concepts such as potentials between information particles and information fields. Using physical concepts such as ligand field theory or band theory as analogies, quantitatively understanding the mechanisms of information diffusion and aggregation could be key to predicting the emergence of new trends and streams of opinion.

## 6. New Paradigms in Understanding Digital Space

The application of first-principles calculations contributes to the construction of new theoretical frameworks for analyzing the flow of information in digital space. In this approach, information is treated as quantized "digital particles," and the study focuses on how information diffuses and aggregates through the complex interaction network formed by these particles. In this theoretical framework, concepts used in physics, such as exchange correlation functions between information particles, the topology of information fields, and even phenomena like information phase transitions and critical phenomena, are applied to formulate the dynamics of information in digital space. Such new perspectives may bring about new theoretical insights in fields such as information science, sociology, and behavioral economics and may function as decision support tools in designing digital media and policymaking. Understanding the fundamental laws of information flow could elucidate and manage phenomena in digital space previously thought to be unpredictable, based on principles of physics.

By harnessing the computational power of these quantum mechanical tools, we aim to provide a comprehensive understanding of how these 2D materials can be leveraged to enhance digital information control and processing. Furthermore, this study explores the potential applications of these materials in the simulation of digital information flow, control strategies, and the mitigation of phenomena such as information bubbles, thereby contributing to the advancement of digital health and the robustness of digital infrastructures. Through this interdisciplinary approach, we bridge the gap between quantum physics and information science, offering novel insights and methodologies that could pave the way for innovative digital technologies.

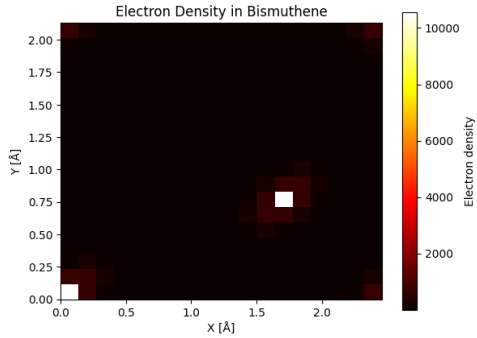

Fig. 3: Electron Density in Bismuthene(Bismuth-Molecule)

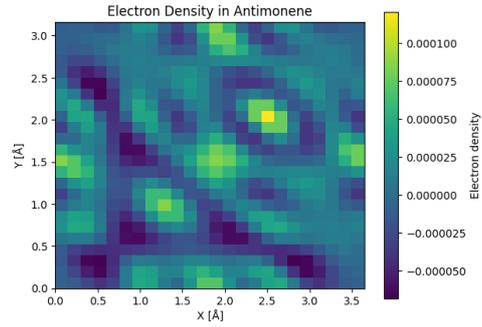

Fig. 4: Electron Density in Antimonene

In the rapidly evolving field of information technology, the exploration of novel materials with unique electronic and optical properties has become increasingly crucial for the development of next-generation digital devices and systems. Among these materials, two-dimensional (2D) honeycomb structures, such as antimonene, tellurene, silane, monosilane, silicene, and bismuthene, have garnered significant attention due to their remarkable characteristics that stem from their quantum mechanical nature. This paper delves into the first-principles calculations based on the Kohn-Sham equations within the framework of density functional theory (DFT) to elucidate the electronic structure and exchange-correlation potential approximations of these materials. By harnessing the computational power of these quantum mechanical tools, we aim to provide a comprehensive understanding of how these 2D materials can be leveraged to enhance digital information control and processing. Furthermore, this study explores the potential applications of these materials in the simulation of digital information flow, control strategies, and the mitigation of phenomena such as information bubbles, thereby contributing to the advancement of digital health and the robustness of digital infrastructures. Through this interdisciplinary approach, we bridge the gap between quantum physics and information science, offering novel insights and methodologies that could pave the way for innovative digital technologies.

Building on the foundational principles outlined in the introduction, this paper ventures into the intricate relationship between the quantum mechanical properties of two-dimensional (2D) materials and their potential applications in digital information systems. The heart of our theoretical investigation lies in the utilization of the Kohn-Sham equations, a cornerstone of density functional theory (DFT), which provides a pragmatic approach to solving many-body problems within quantum mechanics.

The unique honeycomb lattice structures of antimonene, tellurene, silane, monosilane, silicene, and bismuthene offer a playground for electrons that is vastly different from traditional materials. These structures host a variety of electronic phenomena, including unique band structures, quantum confinement, and significant spin-orbit coupling effects, which are pivotal in dictating the materials' electronic and optical behaviors.

At the core of our analysis is the exchange-correlation potential, a critical component in DFT calculations that encapsulates the complex many-body interactions within a quantum system. By approximating this potential, we can extract valuable insights into the electronic density distributions within these materials, offering a window into their intrinsic properties that are paramount for digital information control.

Moreover, the theoretical exploration extends to simulating the behavior of digital information flow within networks structured analogously to the electron flow in these 2D materials. By drawing parallels between the vortex formations in electronic densities and information flow dynamics, we propose models that can mimic and potentially predict the emergence and dissolution of information bubbles within digital networks.

This interdisciplinary approach not only deepens our understanding of the quantum mechanical underpinnings of novel 2D materials but also opens new avenues for leveraging these materials in enhancing digital information infrastructure. The theoretical insights gleaned from this study could lead to the development of more efficient, secure, and resilient digital communication systems, marking a significant step forward in the convergence of quantum materials science and information technology.

The forefront of contemporary material science and theoretical physics is marked by the exploration of two-dimensional (2D) materials, characterized by honeycomb lattice structures. This paper presents a comprehensive analysis of such materials, including Stannene, Stannene, si-nanowires, antimonene, tellurene, silicene, silane, monosilane, and bismuthene, using first-principles calculations grounded in the Kohn-Sham equations of Density Functional Theory (DFT). The focus is on understanding the formation of electronic vortices within these materials and their implications for digital information control and simulation.

The landscape of material science and computational physics is rapidly evolving with the exploration of two-dimensional (2D) materials, particularly those with honeycomb lattice structures like plumbene. This paper aims to delve into the intricate world of such materials by leveraging first-principles calculations, focusing on the Kohn-Sham equations within the framework of Density Functional Theory (DFT). Our primary interest lies in understanding the phenomenon of vortex formation in these materials and its implications for simulations of digital information control.

Two-dimensional materials, epitomized by graphene and extended to others like plumbene, have garnered significant interest due to their exceptional electronic and mechanical properties. These properties are often a direct consequence of the materials' unique lattice structures and the quantum mechanical behavior of electrons within them. Of particular interest in this realm of study is the formation of electronic vortices localized circular movements of electrons that carry significant quantum information and topological characteristics.

The concept of vortices in electronic systems is not just limited to superconductors but extends to normal conducting materials, especially at the nano-scale. In honeycomb lattices, such as those found in plumbene, the interplay between the geometric constraints of the lattice, electron-electron interactions, and external perturbations like electric or magnetic fields can lead to the spontaneous formation of electronic vortices. These vortices, characterized by a whirlpool-like circulation of electron density, are of paramount importance due to their potential to influence the material's electronic properties and, by extension, its conductivity and topological states.

The manipulation and control of electronic vortices in 2D materials open up innovative avenues for digital information processing, potentially revolutionizing how data is stored, processed, and transmitted. This paper explores the theoretical underpinnings of utilizing vortices in 2D materials for digital information control. By encoding data within the topological configurations of electronic vortices, we can envisage new forms of quantum computing architectures and highly secure data storage methods that leverage the quantum mechanical nature of these materials.

This research employs the Kohn-Sham equations of DFT to simulate and analyze the electronic structure of 2D honeycomb materials like plumbene, with a particular focus on understanding vortex formation under various conditions. Despite the promising prospects, several challenges need addressing, including the accurate modeling of these complex quantum phenomena, ensuring the stability of vortex configurations, and their integration into practical computational devices.

In summary, this paper presents a theoretical exploration of the vortex dynamics in 2D honeycomb lattice materials and their potential applications in the realm of digital information control. By harnessing the quantum mechanical properties of these materials, we aim to shed light on novel computational paradigms that could define the future of technology.

# 7. Application of Analogy from Ligand Field Theory and Band Theory in Overcoming Undefined Patterns and Predictive Difficulty

The application of analogies from ligand field theory and band theory in overcoming undefined patterns and predictive difficulty is an intriguing approach to understanding the mechanisms of information diffusion and aggregation. Below, we provide theoretical explanations and concrete examples.

## 7.1 Analogy from Ligand Field Theory

In ligand field theory, we consider how the electronic configuration of atoms or ions is influenced by the surrounding ligand field. When applying this theory to the diffusion of information, each information particle in digital space (e.g., news articles, social media posts) can be likened to atoms or ions, and we analyze how the "energy levels" of these information particles (influence or diffusion rate) change due to the digital environment (ligand field) they exist in.

### 7.1.1 Formulas and Calculation Procedure:

1. Let $E_i$ represent the energy level of information particle $i$, and denote the influence of the ligand field (digital environment) as $V_{\text{env}}$.

$$E'_i = E_i + V_{\text{env}}$$

2. $V_{\text{env}}$ depends on the specific digital environment where the information particle is placed (platform, user group, time period, etc.).

3. Based on the change in energy levels of information particles, evaluate the possibilities of diffusion and aggregation.

$$P_{\text{diff}} = f(E'_i)$$

Here, $P_{\text{diff}}$ is the diffusion probability of information particles, and $f$ is a function dependent on the energy levels.

### 7.2 Analogy from Band Theory

Band theory analyzes the formation of energy bands for electrons within solids. In the analogy of band theory to information diffusion, we understand how information is aggregated and diffused through the "information bands" formed by information particles.

1. Form the energy band for the population of information particles. The width and position of the energy band are determined by interactions between information particles.

$$E_{\text{band}} = E_{\text{base}} + \Delta E_{\text{interaction}}$$

2. Calculate the occupancy probability of energy states within the information band.

$$P_{\text{state}}(E) = \frac{1}{e^{\frac{E-E_F}{kT}} + 1}$$

Here, $E_F$ is the Fermi energy, $k$ is the Boltzmann constant, and $T$ represents the "temperature" of the system (activity or energy).

3. Analyze the mechanisms of information diffusion and aggregation using the energy band and occupancy probability.

By employing these analogies, it becomes possible to quantitatively understand the behavior of information in digital space from a physical perspective and elucidate the principles behind undefined patterns and unpredicted phenomena.

## 8. Commonalities and Differences in the Study of Two-Dimensional Materials with Honeycomb Lattice Structures

In the study of two-dimensional materials with honeycomb lattice structures such as graphene, Stannene, stanane, antimonene, tellurene, and silicene, based on first-principles calculations, particularly those using the Kohn-Sham equation to approximate exchange-correlation potentials, several commonalities and differences can be observed.

### 8.1 Commonalities

(1) **Honeycomb Lattice Structure**: All these materials possess a honeycomb lattice structure, resulting in unique electronic structures and physical properties. Particularly, they are characterized by a linear energy dispersion known as the Dirac cone, which describes the behavior of electrons at low energies.

(2) **Utilization of Kohn-Sham Equation**: By solving the Kohn-Sham equation within density functional theory (DFT), the electronic structures of these materials are calculated. This equation enables the effective reduction of many-body problems to single-electron problems.

(3) **Approximation of Exchange-Correlation Potential**: Proper choice of the exchange-correlation potential approximation is crucial in the calculations of these materials. Typically, local density approximation (LDA) or generalized gradient approximation (GGA) are used, although hybrid functionals or other advanced approximations may be chosen depending on the material properties.

### 8.2 Differences

(1) **Elements and Electronic Structures**: Each material is composed of different elements, leading to significant differences in their electronic structures. For instance, graphene is comprised of carbon atoms, while Stannene is composed of tin atoms. These differences in elements directly affect fundamental physical properties such as band gap, electron density, and optical properties.

(2) **Effect of Spin-Orbit Coupling**: Materials containing heavy elements (e.g., tin atoms in Stannene) exhibit significant effects of spin-orbit coupling, which has a pronounced influence on electronic structures and physical properties. This effect may be negligible in materials containing lighter elements (e.g., carbon atoms in graphene).

(3) **Presence of Band Gap**: Graphene is a zero-band gap semiconductor, whereas other two-dimensional materials like Stannene or antimonene may possess finite band gaps. The presence or absence of a band gap significantly affects the electronic properties and applications of the material.

(4) **Influence of External Conditions**: Interlayer interactions, external pressure, electric fields, and other external conditions may have different effects on the electronic structures of these two-dimensional materials. The behavior of the materials under these conditions can be studied in detail through first-principles calculations.

Understanding these commonalities and differences is crucial for discovering new properties of two-dimensional materials and exploring their applications in nanoscale devices and next-generation electronics.

# 9. Explanation of Commonalities and Differences in Determining Vortex in Twisted Bilayer Structures of Two-Dimensional Materials (e.g., Graphene, Stannene, Stanane, Antimonene, Tellurene, Silicane) Using the Kohn-Sham Equation in Density Functional Theory (DFT)

In this explanation, we will discuss the commonalities and differences in determining vortex in twisted bilayer structures of two-dimensional materials (e.g., graphene, Stannene, stanane, antimonene, tellurene, silicane) using the Kohn-Sham equation in density functional theory (DFT).

## 9.1 Commonalities

(1) **Utilization of the Kohn-Sham Equation**: In all two-dimensional materials, the Kohn-Sham equation is employed to calculate the ground state energy and wave functions of electrons. To analyze the characteristics of the vortex, it is necessary to accurately calculate the electron density distribution and energy levels of these materials.

(2) **Significance of Twisting Angle**: The relative angle between the two layers in twisted bilayer structures significantly affects the electronic structure and consequently influences the formation of the vortex. Changes in electron states due to the angle impact the characteristics of the vortex.

(3) **Approximation of Exchange-Correlation Potential**: The selection of exchange-correlation potentials models the interaction of electrons in the Kohn-Sham equation. While standard approximations such as LDA or GGA are commonly used, appropriate approximations need to be chosen depending on the specific material or situation.

## 9.2 Differences

(1) **Material-Specific Properties**: The unique electronic structures and physical properties of each material significantly influence the formation and characteristics of the vortex. For example, in comparison to graphene, materials containing heavy elements like Stannene or antimonene may exhibit significant effects of spin-orbit coupling, which could impact the behavior of the vortex.

(2) **Impact of Twisting Angle**: In some materials, electron correlations become significant at specific twisting angles, affecting the mechanism of vortex formation. This angle dependence varies depending on the material.

(3) **Band Gap and Metallic State**: Depending on the material, the band gap may open or close at certain twisting angles, while in other materials, a metallic state may persist. These properties directly influence the characteristics of the vortex.

(4) **Need for Advanced Exchange-Correlation Potentials**: In some materials, the use of advanced exchange-correlation potentials may be necessary for an accurate description of the vortex. This is especially true in cases where strong electron correlations or non-local exchange-correlation effects are important.

By considering these commonalities and differences, a precise understanding of the characteristics of vortex in twisted bilayer structures of each two-dimensional material can be achieved, paving the way for the discovery of new physical phenomena and applications.

# 10. Proposal for First Principles Calculations of Vortex in Two-Dimensional Materials with Honeycomb Lattice Structures (e.g., Stannene, Stanane, Si-nanowire, Antimonene, Tellurene, Silicane) in Modeling Information Flow Mimicking the Filter Bubble Phenomenon

In the model of information flow mimicking the filter bubble phenomenon, considering first principles calculations of vortex in two-dimensional materials with honeycomb lattice structures (e.g., Stannene, stanane, Si-nanowire, antimonene, tellurene, silicane), the following ideas and computational processes are proposed.

## 10.1 Computational Ideas

(1) **Modeling the Filter Bubble**:

Represent filter bubbles within the model of information flow as the formation of vortex. In this analogy, homogenization and localization of information are understood as swirling movements of electrons within the material.

(2) **First Principles Calculation Framework**:

Use the Kohn-Sham equation based on density functional theory (DFT) to compute the behavior and interactions of electrons within the two-dimensional material. This determines whether vortex formation occurs under specific conditions.

(3) **Consideration of External Parameters**:

Model external influences on information flow (e.g., reliability of news sources, user preferences) as changes in external parameters such as external electric fields or twisting angles in the first principles calculations. Investigate how these parameter variations affect vortex formation.

### 10.2 Computational Process

(1) **Definition and Optimization of Structure**:

Define the honeycomb lattice structure of each two-dimensional material and perform structure optimization using DFT calculations. This allows for calculations based on actual material properties.

(2) **Calculation of Electron Density**:

Solve the Kohn-Sham equation to calculate the electron density distribution. This step focuses particularly on electron states relevant to vortex formation.

$$\left(-\frac{\hbar^2}{2m}\nabla^2 + V_{\text{eff}}(\mathbf{r})\right)\psi_i(\mathbf{r}) = \varepsilon_i\psi_i(\mathbf{r})$$

(3) **Variation of External Parameters and Their Effects**:

Mimic external conditions in information flow (factors contributing to filter bubble formation) by varying external electric field strength, direction, and twisting angles to calculate their effects on electron density and vortex formation.

(4) **Identification of Vortex Formation Conditions**:

Identify conditions for vortex formation from the electron density distribution. This requires advanced data analysis techniques to identify localized swirling patterns in electron density.

(5) **Application to Information Flow**:

Apply the identified conditions for vortex formation to the model of information flow to simulate the processes of filter bubble formation, maintenance, and disruption. This may provide insights into information management strategies to prevent filter bubbles.

This approach proposes novel ways to understand and control the dynamics of complex information systems by applying principles from physics.

## 11. Detailed Formulas and Computational Process for First Principles Calculations Modeling Vortex Formation as Filter Bubble Phenomenon in Electron Behavior and Interactions within Two-Dimensional Materials

### 11.1 Formulas and Computational Process

(1) **Kohn-Sham Equation**: Based on density functional theory (DFT), the Kohn-Sham equation is represented as follows:

$$\left(-\frac{\hbar^2}{2m}\nabla^2 + V_{\text{eff}}(\mathbf{r})\right)\psi_i(\mathbf{r}) = \varepsilon_i\psi_i(\mathbf{r})$$

where $\psi_i(\mathbf{r})$ represents the $i$-th Kohn-Sham orbital, $\varepsilon_i$ is the corresponding eigenvalue (energy level), and $V_{\text{eff}}(\mathbf{r})$ denotes the effective potential.

(2) **Effective Potential**: The effective potential $V_{\text{eff}}(\mathbf{r})$ consists of the sum of external potential, Hartree term, and exchange-correlation term:

$$V_{\text{eff}}(\mathbf{r}) = V_{\text{ext}}(\mathbf{r}) + \int \frac{n(\mathbf{r}')}{|\mathbf{rr}'|}d\mathbf{r}' + V_{\text{xc}}[n(\mathbf{r})]$$

where $V_{\text{ext}}(\mathbf{r})$ is the external potential due to atomic nuclei, and $V_{\text{xc}}[n(\mathbf{r})]$ represents the exchange-correlation potential.

(3) **Calculation of Electron Density**: Electron density $n(\mathbf{r})$ is computed from the Kohn-Sham orbitals:

$$n(\mathbf{r}) = \sum_i |\psi_i(\mathbf{r})|^2$$

(4) **Identification of Vortex Formation Conditions**: To identify localized swirling patterns (vortex) in the electron density distribution, analyze electron density gradients and curl (rotation):

$$\nabla \times \mathbf{J}(\mathbf{r}) = \nabla \times (n(\mathbf{r})\mathbf{v}(\mathbf{r}))$$

Here, $\mathbf{J}(\mathbf{r})$ represents the current density, and $\mathbf{v}(\mathbf{r})$ is the velocity field of electrons. Vortex corresponds to regions where $\nabla \times \mathbf{J}(\mathbf{r})$ is nonzero.

(5) **Variation of External Conditions**: Analyze how changes in external conditions, such as external electric fields or twisting angles, mimic contributions to filter bubble formation and affect vortex formation.

## 12. 1. Setting Initial Structure and External Conditions

The initial structure defines the honeycomb lattice structure of the target two-dimensional material. External conditions, such as external electric fields or twist angles, are set. At this stage, no specific equations are used, but these conditions are input into the software as parameters for the calculation.

## 13. DFT Calculation and Solution of the Kohn-Sham Equation

The Kohn-Sham equation is expressed in the following form:

$$\left(-\frac{\hbar^2}{2m}\nabla^2 + V_{\text{eff}}(\mathbf{r})\right)\psi_i(\mathbf{r}) = \varepsilon_i \psi_i(\mathbf{r})$$

Here, $V_{\text{eff}}(\mathbf{r})$ is the effective potential, defined as follows:

$$V_{\text{eff}}(\mathbf{r}) = V_{\text{ext}}(\mathbf{r}) + \int \frac{n(\mathbf{r}')}{|\mathbf{rr}'|} d\mathbf{r}' + V_{\text{xc}}[n(\mathbf{r})]$$

Solving the Kohn-Sham equation yields the wave functions $\psi_i(\mathbf{r})$ and energies $\varepsilon_i$ for each electronic state.

## 14. Calculation of Electron Density and Current Density

The electron density is calculated from the wave functions as follows:

$$n(\mathbf{r}) = \sum_i |\psi_i(\mathbf{r})|^2$$

The current density $\mathbf{J}(\mathbf{r})$ is computed from the electron density and the velocity field $\mathbf{v}(\mathbf{r})$ of the electrons:

$$\mathbf{J}(\mathbf{r}) = n(\mathbf{r})\mathbf{v}(\mathbf{r})$$

The velocity field of the electrons is derived from the phase gradients of the wave functions.

## 15. Evaluation of Vortex Formation Conditions

Vortex formation conditions are assessed by the rotation (curl) of the current density:

$$\nabla \times \mathbf{J}(\mathbf{r}) \neq 0$$

Where this condition is met, it can be interpreted that electron flow is forming a vortex.

## 16. Analysis of the Impact of External Condition Changes

The changes in vortex formation conditions due to variations in external conditions (e.g., external electric fields or twist angles) are analyzed. In this process, the conditions are modified, and the Kohn-Sham equation is solved again to compute the electron density and current density under the modified conditions. Based on this, the formation and alteration of vortices are evaluated.

Through these computational processes, changes in electron behavior within two-dimensional materials under specific conditions can be understood and applied as analogies to information-theoretic phenomena such as filter bubbles.

## 17. Modeling External Influence on Information Flow using First Principles Calculations

In modeling the filter bubble phenomenon, simulating the external influence on information flow, such as the reliability of news sources or user preferences, using first principles calculations involves the following steps. Here, we use an analogy that maps external effects in the physical model to changes in external electric fields or twist angles.

### 17.1 Definition and Mapping of Parameters

External parameters, such as the reliability $R$ of news sources and user preferences $P$, are mapped to external electric fields $E$ and twist angles $\theta$, respectively.

Mapping of reliability $R$ to external electric field $E$:
$$E = \alpha \cdot R \quad (1)$$

Mapping of user preferences $P$ to twist angle $\theta$:
$$\theta = \beta \cdot P \quad (2)$$

Here, $\alpha$ and $\beta$ are appropriate scaling factors.

### 17.2 Incorporation of External Parameters into the Kohn-Sham Equation

External parameters are incorporated into the Kohn-Sham equation as follows:

$$\left(-\frac{\hbar^2}{2m}\nabla^2 + V_{\text{eff}}(\mathbf{r}, E, \theta)\right)\psi_i(\mathbf{r}) = \varepsilon_i \psi_i(\mathbf{r})$$

Here, $V_{\text{eff}}(\mathbf{r}, E, \theta)$ is the effective potential considering external parameters.

### 17.3 Calculation of Effective Potential

The effective potential considering external parameters is calculated as follows:

$$V_{\text{eff}}(\mathbf{r}, E, \theta) = V_{\text{ext}}(\mathbf{r}, E) + \int \frac{n(\mathbf{r}')}{|\mathbf{rr}'|} d\mathbf{r}' + V_{\text{xc}}[n(\mathbf{r}), \theta]$$

Here, $V_{\text{ext}}(\mathbf{r}, E)$ represents the external potential accounting for the external electric field $E$, and $V_{\text{xc}}[n(\mathbf{r}), \theta]$ is the exchange-correlation potential considering the twist angle $\theta$.

## 17.4 Calculation of Electron Density and Modeling of Information Flow

Compute the electron density $n(\mathbf{r})$ and interpret it as a model of information flow. Changes in electron density represent the diffusion and accumulation of information.

## 17.5 Analysis of the Impact of Changes in External Parameters on Information Flow

Analyze how changes in the reliability $R$ of news sources and user preferences $P$ affect information flow. Evaluate how information from highly reliable sources or matching user preferences diffuses and accumulates.

Through this approach, the influence of external factors on information flow can be modeled and analyzed from a physical perspective, providing new insights into the formation of filter bubbles and biases in information dissemination.

# 18. Calculation and Analysis of Electron Density and Information Flow

To compute the electron density and model information flow, as well as analyze the impact of changes in external parameters on information flow, the following equations and computational steps are utilized:

## 18.1 Calculation of Electron Density

The electron density $n(\mathbf{r})$ is computed using the Kohn-Sham orbitals $\psi_i(\mathbf{r})$ as follows:

$$n(\mathbf{r}) = \sum_i^{N_{\text{occ}}} |\psi_i(\mathbf{r})|^2$$

Here, $N_{\text{occ}}$ is the number of occupied states.

## 18.2 Modeling Information Flow

Information flow is represented by the electron density, and the diffusion and accumulation of information are modeled as changes in the electron density. The diffusion velocity $v_{\text{info}}$ is determined from the spatial gradient of the electron density:

$$v_{\text{info}}(\mathbf{r}) = -D\nabla n(\mathbf{r})$$

Here, $D$ is the diffusion coefficient.

## 18.3 Analysis of External Parameter Effects

Changes in external parameters are introduced in the form of electric field $E$ and twist angle $\theta$. These parameters influence the effective potential $V_{\text{eff}}$, thereby affecting the electron density and information flow.

The effective potential is modified as follows:

$$V_{\text{eff}}(\mathbf{r}, E, \theta) = V_{\text{ext}}(\mathbf{r}, E) + \int \frac{n(\mathbf{r}')}{|\mathbf{rr}'|} d\mathbf{r}' + V_{\text{xc}}[n(\mathbf{r}), \theta]$$

Here, $V_{\text{ext}}(\mathbf{r}, E)$ represents the external potential accounting for the electric field $E$, and $V_{\text{xc}}[n(\mathbf{r}), \theta]$ is the exchange-correlation potential considering the twist angle $\theta$.

Simulations of electron density and information flow are conducted for different values of external parameters $E$ and $\theta$, and the changes in diffusion patterns are observed. In particular, the diffusion and accumulation of information from highly reliable sources or matching user preferences are evaluated.

This approach allows for a quantitative analysis of the impact of external influences on information flow. It enables a better understanding of phenomena such as filter bubble formation and information bias from a physical perspective.

# 19. Calculation Procedure for Analyzing Vortex Formation and Its Variations

To analyze vortex formation conditions and evaluate changes in vortex formation due to external conditions (such as external electric fields or twist angles), the following computational steps are followed:

## 19.1 Setting up the Kohn-Sham Equation

The Kohn-Sham equation is given by:

$$\left[ -\frac{\hbar^2}{2m}\nabla^2 + V_{\text{eff}}(\mathbf{r}, E, \theta) \right] \psi_i(\mathbf{r}) = \varepsilon_i \psi_i(\mathbf{r})$$

Here, $V_{\text{eff}}(\mathbf{r}, E, \theta)$ is the effective potential considering the external conditions $E$ (external electric field) and $\theta$ (twist angle).

## 19.2 Calculation of Effective Potential

The effective potential $V_{\text{eff}}$ is set up to include the influence of the external electric field $E$ and twist angle $\theta$:

$$V_{\text{eff}}(\mathbf{r}, E, \theta) = V_{\text{ext}}(\mathbf{r}, E) + \int \frac{n(\mathbf{r}')}{|\mathbf{rr}'|} d\mathbf{r}' + V_{\text{xc}}[n(\mathbf{r}), \theta]$$

## 19.3 Calculation of Electron Density and Current Density

The electron density $n(\mathbf{r})$ and current density $\mathbf{j}(\mathbf{r})$ are computed as follows:

Electron density:

$$n(\mathbf{r}) = \sum_i^{N_{\text{occ}}} |\psi_i(\mathbf{r})|^2$$

Current density:

$$\mathbf{j}(\mathbf{r}) = -\frac{i\hbar}{2m} \sum_i^{N_{\text{occ}}} \left( \psi_i^{(}\mathbf{r})\nabla\psi_i(\mathbf{r}) \psi_i(\mathbf{r})\nabla\psi_i^{(}\mathbf{r}) \right)$$

## 19.4 Calculation of Vorticity

Vorticity $\omega(\mathbf{r})$ is obtained from the curl of the current density:

$$\omega(\mathbf{r}) = \nabla \times \mathbf{j}(\mathbf{r})$$

## 19.5 Analysis of Vorticity

The spatial distribution of vorticity $\omega(\mathbf{r})$ is analyzed to evaluate vortex formation under specific conditions. Vorticity peaks or troughs are observed at the center of a vortex, surrounded by swirling current patterns.

## 19.6 Evaluation of Impact of External Condition Changes

By repeating these analysis for different values of external conditions $E$ and $\theta$, the conditions for vortex formation and their variations are evaluated. Particularly, the analysis focuses on how changes in the strength or direction of the external electric field and the magnitude of the twist angle affect vortex formation.

Through this computational process, a quantitative understanding of the influence of external condition changes on vortex formation can be obtained, providing fundamental insights for device design and information flow control applications.

## 20. Modeling Filter Bubble Phenomenon Based on Stannene Characteristics and First Principles Calculations

In modeling the filter bubble phenomenon based on the characteristics of Stannene and performing first principles calculations, the detailed equations and computational process are described as follows:

### 20.1 Stannene-Inspired Kohn-Sham Equation:

Considering the strong spin-orbit coupling in Stannene, the Kohn-Sham equation is represented as follows:

$$\left(-\frac{\hbar^2}{2m}\nabla^2 + V_{\text{eff}}(\mathbf{r}) + \xi \cdot \mathbf{L} \cdot \mathbf{S}\right)\psi_i(\mathbf{r}) = \varepsilon_i \psi_i(\mathbf{r})$$

Here, $\xi$ denotes the strength of spin-orbit coupling, $\mathbf{L}$ represents the orbital angular momentum, and $\mathbf{S}$ denotes the spin angular momentum. The inclusion of the spin-orbit term significantly affects the electronic structure of Stannene.

### 20.2 Effective Potential:

For Stannene, the effective potential $V_{\text{eff}}(\mathbf{r})$ may be modified to account for the influence of spin-orbit coupling:

$$V_{\text{eff}}(\mathbf{r}) = V_{\text{ext}}(\mathbf{r}) + \int \frac{n(\mathbf{r}')}{|\mathbf{rr}'|}d\mathbf{r}' + V_{\text{xc}}[n(\mathbf{r})] + V_{\text{SO}}$$

Here, $V_{\text{SO}}$ represents an additional potential term due to spin-orbit interaction.

### 20.3 Electron Density and Vortex Formation Criteria:

In evaluating the electron density and vortex formation criteria in Stannene, the influence of spin-orbit coupling must be considered. The electron density is calculated using the following expression:

$$n(\mathbf{r}) = \sum_{i,\sigma} |\psi_{i\sigma}(\mathbf{r})|^2$$

Here, $\sigma$ denotes the spin index (spin-up or spin-down).

The criteria for vortex formation involve identifying vortex-like patterns in the electron density distribution. Considering spin-orbit coupling, the spin-dependent current density $\mathbf{J}_\sigma(\mathbf{r})$ plays a crucial role in the analysis of vortex formation criteria.

### 20.4 Effects of External Conditions:

The influence of changes in external electric fields or twist angles on the electronic structure in Stannene interacts significantly with spin-orbit coupling and requires special attention. Changes in electron density and vortex formation criteria due to variations in external conditions are analyzed through the following steps:

1. Modify the external conditions and recalculate the Kohn-Sham equation. 2. Reevaluate the electron density and current density under the modified conditions and observe the formation and changes of vortices.

Through this approach, a deeper understanding of the flow and localization mechanisms of information in the filter bubble phenomenon can be obtained, potentially leading to novel approaches for information management and diffusion.

## 21. Detailed Formulas and Calculation Procedure for Modeling the Filter Bubble Phenomenon Considering the Characteristics of Si-Nanowires

### 21.1 Setting the Structure and Initial Conditions of Si-Nanowires

The initial structure of Si-nanowires is defined based on crystallographic parameters. Typically, a vertically elongated honeycomb structure is employed. Additionally, external conditions such as electric field $E$ and twist angle $\theta$ are set.

### 21.2 Adjustment of Effective Potential

The effective potential $V_{\text{eff}}(\mathbf{r})$ is adjusted considering the influence of external electric fields and twist angles. This includes adding potentials due to the external electric field $V_E(\mathbf{r}, E)$ and the twist angle $V_\theta(\mathbf{r}, \theta)$.

## 21.3 Calculation of Electron Density and Current Density

Solving the Kohn-Sham equations yields electron wave functions $\psi_i(\mathbf{r})$, from which the electron density $n(\mathbf{r})$ is computed. Furthermore, using the electron velocity field $\mathbf{v}(\mathbf{r})$, the current density $\mathbf{J}(\mathbf{r})$ is calculated.

## 21.4 Analysis of the Impact of External Conditions

Changes in the electron density and current density due to variations in the external electric field $E$ and twist angle $\theta$ are computed. From these changes, the conditions for vortex formation are reassessed, specifically identifying regions where $\nabla \times \Delta \mathbf{J}(\mathbf{r}; E, \theta) \neq 0$.

Through this computational process, a detailed analysis of the influence of external condition changes on electron behavior is conducted, contributing to a deeper understanding of the modeling of the filter bubble phenomenon considering the characteristics of Si-nanowires.

article amsmath

## 22. Detailed Formulas and Calculation Procedure for Modeling the Filter Bubble Phenomenon Considering the Characteristics of Antimonene and Tellurene

### 22.1 Characteristics of Antimonene and Tellurene

Antimonene and Tellurene, being composed of heavy elements, exhibit significant spin-orbit coupling effects. As a result, they manifest unique features in electronic structure and optical properties, influencing the conditions for vortex formation.

### 22.2 Formulas and Calculation Process

1. Application of the Kohn-Sham Equation: The Kohn-Sham equation accounting for spin-orbit interaction is established based on density functional theory (DFT).

$$\left(-\frac{\hbar^2}{2m}\nabla^2 + V_{\text{eff}}(\mathbf{r}) + \xi(\mathbf{r})\mathbf{L} \cdot \mathbf{S}\right)\psi_i(\mathbf{r}) = \varepsilon_i \psi_i(\mathbf{r})$$

Here, $\xi(\mathbf{r})$ represents the strength of position-dependent spin-orbit coupling, and $\mathbf{L}$ and $\mathbf{S}$ denote orbital angular momentum and spin angular momentum, respectively.

2. Calculation of Effective Potential $V_{\text{eff}}$: The effective potential includes external potential, Hartree term, and exchange-correlation potential. Spin-orbit interaction is also considered at this stage.

$$V_{\text{eff}}(\mathbf{r}) = V_{\text{ext}}(\mathbf{r}) + \int \frac{n(\mathbf{r}')}{|\mathbf{r}\mathbf{r}'|} d\mathbf{r}' + V_{\text{xc}}(\mathbf{r})$$

3. Calculation of Electron Density $n(\mathbf{r})$: Electron density considering spin is computed across all occupied Kohn-Sham orbitals.

$$n(\mathbf{r}) = \sum_i f_i |\psi_i(\mathbf{r})|^2$$

Here, $f_i$ represents the occupation numbers.

4. Identification of Vortex Formation Conditions: The current density $\mathbf{J}(\mathbf{r})$ and its curl are computed to analyze the conditions for vortex formation. In the presence of strong spin-orbit interaction, vortex formation may exhibit complex patterns.

$$\nabla \times \mathbf{J}(\mathbf{r}) = \text{indicator of vortex formation}$$

5. Analysis of the Impact of External Conditions: The effects of changes in external electric fields and twist angles on electron density distribution and vortex formation are evaluated. In Antimonene and Tellurene, spin-orbit interaction significantly influences these effects.

### 22.3 First-Principles Calculation Process for Vortex Formation in Antimonene and Tellurene

1. Definition and Optimization of Structure
Structure Definition:

atomic arrangement and lattice constants of the two-dimensional.

Structure Optimization:

$$E_{\text{total}} = \min \left(E_{\text{kinetic}} + E_{\text{potential}}\right)$$

Here, $E_{\text{total}}$ is the total energy, $E_{\text{kinetic}}$ is the kinetic, and $E_{\text{potential}}$ is the

2. Calculation of Electron Density
Kohn-Sham Equation:

$$\left(-\frac{\hbar^2}{2m}\nabla^2 + V_{\text{eff}}(\mathbf{r})\right)\psi_i(\mathbf{r}) = \varepsilon_i \psi_i(\mathbf{r})$$

Effective Potential $V_{\text{eff}}$:

$$V_{\text{eff}}(\mathbf{r}) = V_{\text{ext}}(\mathbf{r}) + V_{\text{Hartree}}(\mathbf{r}) + V_{\text{xc}}(\mathbf{r})$$

Electron Density:

$$n(\mathbf{r}) = \sum_i f_i |\psi_i(\mathbf{r})|^2$$

Here, $f_i$ represents the occupation numbers.

3. Analysis of External Parameter Effects
Application of External Electric Field:

$$V_{\text{ext}}(\mathbf{r}) = V_{\text{nuclei}}(\mathbf{r}) + e\mathbf{E} \cdot \mathbf{r}$$

Here, $\mathbf{E}$ is the external electric field.

Variation in Twist Angle:

4. Identification of Vortex Formation Conditions
Current Density and its Curl:

$$\mathbf{J}(\mathbf{r}) = -e \sum_i f_i \text{Im} \left[ \psi_i^*(\mathbf{r}) \nabla \psi_i(\mathbf{r}) \right]$$

$$\nabla \times \mathbf{J}(\mathbf{r}) = \text{indicator of vortex formation}$$

5. Application to Information Flow
Modeling of Filter Bubble:

Interpret the obtained vortex patterns as homogenization or localization of information flow.

Proposal of Control Strategies: Control vortex formation by adjusting external conditions and simulate the formation or dissolution of filter bubbles.

Through this calculation process, it is possible to identify the vortex formation conditions considering the influence of spin-orbit interaction in Antimonene and Tellurene and apply them to analyze the filter bubble phenomenon in information flow.

# 23. Detailed Formulas and Calculation Procedure for Modeling the Filter Bubble Phenomenon Based on Silicon Characteristics

## 23.1 Kohn-Sham Equation

To analyze the electronic structure of silicon, we utilize the Kohn-Sham equation, which is expressed as follows:

$$\left( -\frac{\hbar^2}{2m} \nabla^2 + V_{\text{eff}}(\mathbf{r}) \right) \psi_i(\mathbf{r}) = \varepsilon_i \psi_i(\mathbf{r})$$

Here, $\psi_i(\mathbf{r})$ represents the $i$-th Kohn-Sham orbital, $\varepsilon_i$ is the corresponding energy eigenvalue, and $V_{\text{eff}}(\mathbf{r})$ is the effective potential.

## 23.2 Effective Potential

The effective potential $V_{\text{eff}}(\mathbf{r})$ consists of the sum of the external potential $V_{\text{ext}}(\mathbf{r})$, Hartree term, and exchange-correlation term $V_{\text{xc}}(\mathbf{r})$.

$$V_{\text{eff}}(\mathbf{r}) = V_{\text{ext}}(\mathbf{r}) + \int \frac{n(\mathbf{r}')}{|\mathbf{r} - \mathbf{r}'|} d\mathbf{r}' + V_{\text{xc}}(\mathbf{r})$$

## 23.3 Calculation of Electron Density

The electron density $n(\mathbf{r})$ is computed through the Kohn-Sham orbitals.

$$n(\mathbf{r}) = \sum_i |\psi_i(\mathbf{r})|^2$$

## 23.4 Identification of Vortex Formation Conditions

To identify the conditions for vortex formation from the distribution of electron density, we compute the current density $\mathbf{J}(\mathbf{r})$ and its curl.

$$\nabla \times \mathbf{J}(\mathbf{r}) = \nabla \times (n(\mathbf{r})\mathbf{v}(\mathbf{r}))$$

Here, $\mathbf{v}(\mathbf{r})$ represents the velocity field of electrons. Vortices are identified as regions where this curl is non-zero.

## 23.5 Variation of External Conditions

To evaluate the influence of changes in external parameters such as external electric fields and twist angles on vortex formation, we vary these parameters and reanalyze the Kohn-Sham equation.

## 23.6 Calculation Procedure

1. Initialization: Set up the initial structure based on the two-dimensional honeycomb structure of silicon. 2. Execution of DFT Calculation: Solve the Kohn-Sham equation to compute the electron density. 3. Calculation of Current Density and Curl: Compute the current density from the electron density and analyze its curl to confirm the presence of vortices. 4. Evaluation of the Impact of External Condition Changes: Vary external parameters such as external electric fields and twist angles and evaluate how they affect vortex formation. 5. Application of Results: Apply the obtained conditions for vortex formation to analyze the filter bubble phenomenon and propose control strategies for information flow.

Through this calculation process, we aim to deepen our understanding of electronic behavior and interactions in silicon, particularly regarding vortex formation, and explore applications in an information-theoretical context.

# Detailed Explanation of Modeling the Filter Bubble Phenomenon Based on Silicon Characteristics

## Kohn-Sham Equation

The Kohn-Sham equation is the core equation used to determine the ground-state energy of electrons and is expressed in the following form:

$$\left( -\frac{\hbar^2}{2m} \nabla^2 + V_{\text{eff}}(\mathbf{r}) \right) \psi_i(\mathbf{r}) = \varepsilon_i \psi_i(\mathbf{r})$$

Here,

$\psi_i(\mathbf{r})$ represents the $i$-th electron wave function,

$\varepsilon_i$ is the corresponding energy eigenvalue, and

$V_{\text{eff}}(\mathbf{r})$ is the effective potential felt by the electron.

### Effective Potential

The effective potential $V_{\text{eff}}(\mathbf{r})$ is defined as the sum of the external potential $V_{\text{ext}}(\mathbf{r})$, the Hartree term, and the exchange-correlation potential $V_{\text{xc}}(\mathbf{r})$.

$$V_{\text{eff}}(\mathbf{r}) = V_{\text{ext}}(\mathbf{r}) + \int \frac{n(\mathbf{r}')}{|\mathbf{r} - \mathbf{r}'|} d\mathbf{r}' + V_{\text{xc}}[n(\mathbf{r})]$$

### Calculation of Electron Density

The electron density $n(\mathbf{r})$ is calculated as the sum of the squares of the absolute values of all electron wave functions.

$$n(\mathbf{r}) = \sum_i |\psi_i(\mathbf{r})|^2$$

### Identification of Vortex Formation Conditions

To identify vortex formation, we analyze the gradients of the electron density and the curl of the current density $\mathbf{J}(\mathbf{r})$. The current density is given by the product of the electron density and the velocity field $\mathbf{v}(\mathbf{r})$.

$$\mathbf{J}(\mathbf{r}) = n(\mathbf{r})\mathbf{v}(\mathbf{r})$$

The curl is computed as follows:

$$\nabla \times \mathbf{J}(\mathbf{r}) = \nabla \times (n(\mathbf{r})\mathbf{v}(\mathbf{r}))$$

Vortices are identified as regions where this curl is non-zero.

### Variation of External Conditions

We analyze how changes in external conditions, such as external electric fields or twist angles, affect electron density, current density, and vortex formation. This involves solving the Kohn-Sham equation again after modifying the conditions.

Through this calculation process, we can understand the behavior and interactions of electrons in silicon and identify the conditions for vortex formation under specific circumstances. This approach can also be extended to modeling information flow and applying it to the filter bubble phenomenon.

## Modeling the Filter Bubble Phenomenon Considering Bismuthene Properties

### Application of Kohn-Sham Equation

In the case of bismuthene, the Kohn-Sham equation is expressed as follows:

$$\left(-\frac{\hbar^2}{2m}\nabla^2 + V_{\text{eff}}(\mathbf{r})\right)\psi_i(\mathbf{r}) = \varepsilon_i \psi_i(\mathbf{r})$$

Here, it's necessary to consider the heavy mass of bismuth atoms and the strong spin-orbit coupling (SOC) effect, which significantly influence the electron structure and band gap. These effects can be properly handled in the calculation of the effective potential $V_{\text{eff}}(\mathbf{r})$, especially by including the SOC term.

### Adjustment of Effective Potential

The effective potential of bismuthene, considering spin-orbit coupling, is expressed as:

$$V_{\text{eff}}(\mathbf{r}) = V_{\text{ext}}(\mathbf{r}) + \int \frac{n(\mathbf{r}')}{|\mathbf{r} - \mathbf{r}'|} d\mathbf{r}' + V_{\text{xc}}[n(\mathbf{r})] + V_{\text{SOC}}$$

Here, $V_{\text{SOC}}$ represents the spin-orbit coupling term, reflecting aspects of the electron structure specific to bismuthene.

### Calculation of Electron Density and Vortex Formation Conditions

The electron density $n(\mathbf{r})$ is calculated as the sum of squares of Kohn-Sham orbitals. Vortex formation conditions are identified by analyzing the local gradients and curls of the electron density. In the case of bismuthene, spin-orbit coupling may play an important role in the characteristics of vortices.

### Analysis of External Conditions Influence

Changes in external conditions, such as external electric fields or twist angles, are expected to significantly affect the electron structure and vortex formation in bismuthene. In particular, the strong effect of spin-orbit coupling may make the system's response to changes in external conditions very complex. To analyze these effects, it's necessary to solve the Kohn-Sham equation again after modifying the external conditions and evaluate the new electron structure and vortex formation conditions.

### Control Strategies for Filter Bubbles and Information Flow

The obtained results can be applied to understand the filter bubble phenomenon and develop control strategies for information flow. Understanding how the specific physical properties of bismuthene influence information flow models may provide new insights into digital information management.

Thus, the first-principles computational approach considering bismuthene properties provides an effective means for modeling information flow mimicking the filter bubble phenomenon, opening up new possibilities for interdisciplinary research at the intersection of physics and information science.

The computational process for analyzing the effects of external conditions on bismuthene follows the steps outlined below:

## Changes in External Conditions and Their Effects

### a. Introduction of External Electric Field

To consider the influence of an external electric field, an electric field term is added to the effective potential $V_{\text{eff}}(\mathbf{r})$. When an electric field $\mathbf{E}$ is present, the potential energy term $V_{\text{ext}}(\mathbf{r})$ is modified to include the field, resulting in the following expression:

$$V_{\text{eff}}(\mathbf{r}) = V_{\text{ext}}(\mathbf{r}) - e\mathbf{E} \cdot \mathbf{r} + \int \frac{n(\mathbf{r}')}{|\mathbf{r} - \mathbf{r}'|} d\mathbf{r}' + V_{\text{xc}}[n(\mathbf{r})]$$

Here, $e$ represents the charge of an electron, and $\mathbf{E} \cdot \mathbf{r}$ denotes the dot product of the electric field and the position vector.

### b. Change in Torsion Angle

When changing the angle in a twisted bilayer structure, the atomic arrangement in each layer is altered. This change is modeled by adjusting the external potential $V_{\text{ext}}(\mathbf{r})$ within the effective potential $V_{\text{eff}}(\mathbf{r})$. Altering the angle may affect the atomic distances and interactions, potentially influencing the overall electron density distribution.

### c. Recalculation of Electron Density and Current Density

After changing the external conditions, the Kohn-Sham equations are solved again to obtain the new electron density $n(\mathbf{r})$:

$$n(\mathbf{r}) = \sum_i |\psi_i(\mathbf{r})|^2$$

Subsequently, the current density $\mathbf{J}(\mathbf{r})$ is recalculated based on the new electron density distribution:

$$\mathbf{J}(\mathbf{r}) = -\frac{i\hbar}{2m} \left( \psi^*(\mathbf{r}) \nabla \psi(\mathbf{r}) - \psi(\mathbf{r}) \nabla \psi^*(\mathbf{r}) \right)$$

### d. Analysis of Vortex Formation Conditions

From the new electron density and current density, the curl $\nabla \times \mathbf{J}(\mathbf{r})$ is computed to analyze changes in the vortex formation conditions. This allows for evaluating the presence and characteristics of vortex formation under modified external conditions.

### e. Analysis of Effects

The obtained results enable the analysis of how changes in external electric fields or torsion angles affect electron behavior and vortex formation within bismuthene. This understanding can be applied to control information flow and model filter bubble phenomena.

Through these computational steps, understanding the impact of changes in external conditions on the electron structure and physical properties of bismuthene can be achieved, with potential applications to problems in information science.

Theoretical discussion on the approximation of exchange-correlation potentials in two-dimensional materials with honeycomb lattice structures such as Stannene, stanane, sinanowire, antimonene, tellurene, silicene, and bismuthene in the context of first-principles calculations is presented.

## Basic Principles

First-principles calculations, particularly density functional theory (DFT), are powerful tools for determining the ground state energy and wave functions of electrons. In DFT, the many-body problem is simplified into an effective one-electron problem, with the Kohn-Sham equations at its core.

$$\left( -\frac{\hbar^2}{2m} \nabla^2 + V_{\text{eff}}(\mathbf{r}) \right) \psi_i(\mathbf{r}) = \varepsilon_i \psi_i(\mathbf{r})$$

Here, $V_{\text{eff}}(\mathbf{r})$ is the effective potential, composed of the external potential, electron-electron Coulomb interaction (Hartree term), and exchange-correlation potential $V_{\text{xc}}$.

## Approximations to Exchange-Correlation Potentials

The exchange-correlation potential represents the exchange and correlation effects between electrons, directly affecting the accuracy of DFT calculations. However, since the exact form is unknown, various approximations have been proposed.

1. **Local Density Approximation (LDA)**: LDA assumes that the exchange-correlation energy at each point depends only on the electron density at that point. This approximation is based on calculations of exchange-correlation energy for a uniform electron gas. While LDA provides surprisingly good results for many systems and is straightforward, its accuracy may decrease in regions of rapid electron density changes (e.g., bonding regions or near the band gap).

2. **Generalized Gradient Approximation (GGA)**: GGA expresses the exchange-correlation energy as a function of both the electron density and its gradient. This allows for a more accurate treatment of local changes in electron density, often yielding improved results compared to LDA. GGA is widely used for structural determination of molecules and solids, as well as for computing reaction barriers.

3. **Hybrid Functionals**: Hybrid functionals utilize a combination of local and non-local exchange energies, typically including a fraction of exact exchange energy from Hartree-Fock theory. Hybrid functionals are known to provide more accurate results, especially for systems with strong electron correlation or in calculations involving electronic excited states, but they come with the drawback of higher computational cost.

## Applications in Two-Dimensional Materials

The choice of exchange-correlation potential approximation in two-dimensional materials such as Stannene, stanane, Si-nanowire, antimonene, tellurene, silicene, and bismuthene depends on the characteristics of the material under study and the objectives of the research. For example, hybrid functionals might be preferred when precise prediction of band gaps or electronic states is required, while computationally efficient LDA or GGA may be chosen for studies of large systems or complex heterostructures.

In the study of Stannene and other two-dimensional materials, the approximation of exchange-correlation potentials is a crucial element of DFT calculations. The choice of appropriate approximation depends greatly on the desired physical properties and computational feasibility, necessitating a balance between accuracy and efficiency.

We will explain the specific equations and procedures for first-principles calculations based on the properties of arsene.

## Kohn-Sham Equation

The Kohn-Sham equation is used to determine the wave functions and energy levels of electrons and is represented as follows:

$$\left(-\frac{\hbar^2}{2m}\nabla^2 + V_{\text{eff}}(\mathbf{r})\right)\psi_i(\mathbf{r}) = \varepsilon_i\psi_i(\mathbf{r})$$

Here,

$\psi_i(\mathbf{r})$ is the wave function of the $i$-th electron,

$\varepsilon_i$ is the energy level of the $i$-th electron, and

$V_{\text{eff}}(\mathbf{r})$ is the effective potential.

## Effective Potential

The effective potential is calculated considering electron-electron interactions and external potentials.

$$V_{\text{eff}}(\mathbf{r}) = V_{\text{ext}}(\mathbf{r}) + \int \frac{n(\mathbf{r}')}{|\mathbf{r} - \mathbf{r}'|}d\mathbf{r}' + V_{\text{xc}}[n(\mathbf{r})]$$

$V_{\text{ext}}(\mathbf{r})$ is the external potential (e.g., Coulomb potential from nuclei).

The second term represents electron-electron interactions (Hartree term).

$V_{\text{xc}}[n(\mathbf{r})]$ is the exchange-correlation potential depending on the electron density $n(\mathbf{r})$.

## Calculation of Electron Density

The electron density is calculated from Kohn-Sham orbitals.

$$n(\mathbf{r}) = \sum_i |\psi_i(\mathbf{r})|^2$$

## Identification of Vortex Formation Conditions

Identification of localized vortex patterns in the electron density distribution is done by analyzing the gradient and curl (rotation) of the electron density.

$$\mathbf{J}(\mathbf{r}) = -\frac{i\hbar}{2m}(\psi^*\nabla\psi - \psi\nabla\psi^*) - \frac{e^2}{m}\mathbf{A}\psi^*\psi$$

Here, $\mathbf{A}$ is the vector potential. Vortex is identified where the curl of the current density $\nabla \times \mathbf{J}(\mathbf{r})$ is nonzero.

## Variation of External Conditions and Their Effects

Changes in external conditions such as external electric fields or twist angles are simulated to analyze their effects on electron density and vortex formation within arsene. In this step, terms related to external electric fields or twist angles are added to the above equations, and their effects on the system's properties are calculated.

Through these computational processes, detailed analysis of electron behavior, interactions within arsene, and vortex formation under specific conditions can be conducted. The insights gained can be applied to understanding the filter bubble phenomenon and developing strategies for controlling information flow.

We will explain the specific equations and procedures for first-principles calculations based on the properties of arsene.

## Kohn-Sham Equation

The Kohn-Sham equation is used to determine the wave functions and energy levels of electrons and is represented as follows:

$$\left(-\frac{\hbar^2}{2m}\nabla^2 + V_{\text{eff}}(\mathbf{r})\right)\psi_i(\mathbf{r}) = \varepsilon_i\psi_i(\mathbf{r})$$

Here,

$\psi_i(\mathbf{r})$ is the wave function of the $i$-th electron,

$\varepsilon_i$ is the energy level of the $i$-th electron, and

$V_{\text{eff}}(\mathbf{r})$ is the effective potential.

## Effective Potential

The effective potential is calculated considering electron-electron interactions and external potentials.

$$V_{\text{eff}}(\mathbf{r}) = V_{\text{ext}}(\mathbf{r}) + \int \frac{n(\mathbf{r}')}{|\mathbf{r} - \mathbf{r}'|} d\mathbf{r}' + V_{\text{xc}}[n(\mathbf{r})]$$

$V_{\text{ext}}(\mathbf{r})$ is the external potential (e.g., Coulomb potential from nuclei).

The second term represents electron-electron interactions (Hartree term).

$V_{\text{xc}}[n(\mathbf{r})]$ is the exchange-correlation potential depending on the electron density $n(\mathbf{r})$.

## Calculation of Electron Density

The electron density is calculated from Kohn-Sham orbitals.

$$n(\mathbf{r}) = \sum_i |\psi_i(\mathbf{r})|^2$$

## Identification of Vortex Formation Conditions

Identification of localized vortex patterns in the electron density distribution is done by analyzing the gradient and curl (rotation) of the electron density.

$$\mathbf{J}(\mathbf{r}) = -\frac{i\hbar}{2m}(\psi^* \nabla \psi - \psi \nabla \psi^*) - \frac{e^2}{m} \mathbf{A} \psi^* \psi$$

Here, $\mathbf{A}$ is the vector potential. Vortex is identified where the curl of the current density $\nabla \times \mathbf{J}(\mathbf{r})$ is nonzero.

## Variation of External Conditions and Their Effects

Changes in external conditions such as external electric fields or twist angles are simulated to analyze their effects on electron density and vortex formation within arsene. In this step, terms related to external electric fields or twist angles are added to the above equations, and their effects on the system's properties are calculated.

Through these computational processes, detailed analysis of electron behavior, interactions within arsene, and vortex formation under specific conditions can be conducted. The insights gained can be applied to understanding the filter bubble phenomenon and developing strategies for controlling information flow.

We will explain the computational process based on the properties of Monosilane (SiH$_4$), which, although not inherently a two-dimensional material, is sometimes referred to in the study of silicon-based two-dimensional materials.

## Kohn-Sham Equation

In the framework of density functional theory (DFT), the Kohn-Sham equation is used to understand the electronic properties of silicon-based two-dimensional materials:

$$\left(-\frac{\hbar^2}{2m} \nabla^2 + V_{\text{eff}}(\mathbf{r})\right) \psi_i(\mathbf{r}) = \varepsilon_i \psi_i(\mathbf{r})$$

Here, $\psi_i(\mathbf{r})$ represents the $i$-th Kohn-Sham orbital, $\varepsilon_i$ corresponds to its eigenvalue (energy), and $V_{\text{eff}}(\mathbf{r})$ denotes the effective potential.

## Effective Potential

The effective potential is composed of the external potential, Hartree potential (electron-electron interaction), and exchange-correlation potential:

$$V_{\text{eff}}(\mathbf{r}) = V_{\text{ext}}(\mathbf{r}) + \int \frac{n(\mathbf{r}')}{|\mathbf{r} - \mathbf{r}'|} d^3 r' + V_{\text{xc}}[n(\mathbf{r})]$$

## Calculation of Electron Density

The electron density is obtained from the superposition of Kohn-Sham orbitals:

$$n(\mathbf{r}) = \sum_i |\psi_i(\mathbf{r})|^2$$

## Identification of Vortex Formation Conditions

To investigate the presence of localized vortex patterns in the electron density, it is necessary to compute the current density and its curl. The current density is given by:

$$\mathbf{J}(\mathbf{r}) = -\frac{i\hbar}{2m_e}(\psi^* \nabla \psi - \psi \nabla \psi^*)$$

Vortex is indicated by the rotation of the current density, i.e., $\nabla \times \mathbf{J}(\mathbf{r})$.

## Variation of External Conditions and Their Effects

To mimic phenomena similar to filter bubble formation, parameters such as external electric fields or twist angles are varied, and their effects on the system's electron density and vortex formation are investigated. This is treated as changes in external conditions in the model, which may affect the electron density and effective potential.

Through this process, understanding of electron behavior, interactions within Monosilane and related silicon-based two-dimensional materials, and vortex formation under specific conditions can be achieved and applied to the understanding of information science contexts and the phenomenon of filter bubbles.

Here are the specific equations and computational procedures used for performing first-principles calculations of Monosilane and similar two-dimensional materials.

## Kohn-Sham Equation

The Kohn-Sham equation is a fundamental equation used to transform the many-body problem of the electron system into an effective single-particle problem. This equation is represented as:

$$\left(-\frac{\hbar^2}{2m}\nabla^2 + V_{\text{eff}}(\mathbf{r})\right)\psi_i(\mathbf{r}) = \varepsilon_i \psi_i(\mathbf{r})$$

Here, $\psi_i(\mathbf{r})$ represents the $i$-th Kohn-Sham orbital, $\varepsilon_i$ corresponds to the corresponding energy eigenvalue, and $V_{\text{eff}}(\mathbf{r})$ is the effective potential.

## Effective Potential

The effective potential $V_{\text{eff}}(\mathbf{r})$ is given by the sum of the external potential $V_{\text{ext}}(\mathbf{r})$, the Hartree potential due to electron-electron interactions, and the exchange-correlation potential $V_{\text{xc}}[n(\mathbf{r})]$:

$$V_{\text{eff}}(\mathbf{r}) = V_{\text{ext}}(\mathbf{r}) + \int \frac{n(\mathbf{r}')}{|\mathbf{r} - \mathbf{r}'|} d^3 r' + V_{\text{xc}}[n(\mathbf{r})]$$

## Calculation of Electron Density

The electron density is obtained by summing over all occupied Kohn-Sham orbitals:

$$n(\mathbf{r}) = \sum_i |\psi_i(\mathbf{r})|^2$$

## Identification of Vortex Formation Conditions

To identify the conditions for vortex formation, the spatial distribution and gradients of the electron density are analyzed. The current density $\mathbf{J}(\mathbf{r})$ is defined as:

$$\mathbf{J}(\mathbf{r}) = -\frac{i\hbar}{2m_e}(\psi^* \nabla \psi - \psi \nabla \psi^*)$$

Vortex is identified by the rotation of the current density, i.e., $\nabla \times \mathbf{J}(\mathbf{r})$ in non-zero regions.

## Variation of External Conditions and Their Effects

By changing conditions such as external electric fields or twist angles, the effects on the system's electron density and vortex formation are investigated. These variations affect the effective potential, resulting in changes in the electron density distribution.

This computational process forms the basis for understanding the electronic properties and interactions of Monosilane and other two-dimensional materials, which can be applied to information science problems, particularly in modeling and analyzing the filter bubble phenomenon.

To simulate digital information control using the Kohn-Sham equation applied to materials like Bismuth telluride, a unique approach is required to understand the material's electronic properties and use them as metaphors for information transmission and control. Below, I will discuss the theoretical perspectives of this approach.

Bismuth telluride is an intriguing two-dimensional material due to its unique band structure and electronic properties. By solving the Kohn-Sham equation, one can calculate fundamental electronic properties of Bismuth telluride, such as electron density distribution, band gap, and surface states. These properties are crucial for understanding how the material responds to light and electrical signals.

Understanding the electronic properties of materials like Bismuth telluride opens up new opportunities for controlling and transmitting digital information. For example, by altering the material's electronic states, it may be possible to use it as a switch to control the on-off state of information. Additionally, by tuning the band gap, it could function as a selective filter for absorbing or transmitting specific frequencies of light.

Using first-principles calculations based on the Kohn-Sham equation to accurately simulate the electronic properties of Bismuth telluride and interpreting the results in the context of information control is crucial. This process involves evaluating how the material responds to external stimuli (such as light or electric fields) and assessing factors like the efficiency, speed, and stability of information transmission.

This approach bridges the fields of electronic properties and information science, presenting several theoretical challenges. These include accurately predicting material properties applicable to real information transmission systems, modeling the material's response to external stimuli, and understanding the interactions between materials for information transmission. Addressing these challenges requires a multi-disciplinary approach integrating computational physics, materials science, and information science.

In conclusion, simulating digital information control using the Kohn-Sham equation applied to materials like Bismuth telluride holds potential for discovering new materials and developing advanced information processing technologies. However, achieving this requires close collaboration between theoretical insights and experimental validation.

The simulation of digital information control using the Kohn-Sham equation applied to materials like Bismuth telluride requires a unique approach to understand the material's electronic properties and utilize them as metaphors for information transmission and control. Below, we will elaborate on the theoretical outlook of this approach.

Bismuth telluride is an intriguing two-dimensional material due to its unique band structure and electronic properties. By solving the Kohn-Sham equation, one can calculate fundamental electronic properties of Bismuth telluride, such as electron density distribution, band gap, and surface states. These properties are crucial for understanding how the material responds to light and electrical signals.

Understanding the electronic properties of materials like Bismuth telluride opens up new opportunities for controlling and transmitting digital information. For example, by altering the material's electronic states, it may be possible to use it as a switch to control the on-off state of information. Additionally, by tuning the band gap, it could function as a selective filter for absorbing or transmitting specific frequencies of light.

Using first-principles calculations based on the Kohn-Sham equation to accurately simulate the electronic properties of Bismuth telluride and interpreting the results in the context of information control is crucial. This process involves evaluating how the material responds to external stimuli (such as light or electric fields) and assessing factors like the efficiency, speed, and stability of information transmission.

This approach bridges the fields of electronic properties and information science, presenting several theoretical challenges. These include accurately predicting material properties applicable to real information transmission systems, modeling the material's response to external stimuli, and understanding the interactions between materials for information transmission. Addressing these challenges requires a multidisciplinary approach integrating computational physics, materials science, and information science.

In conclusion, simulating digital information control using the Kohn-Sham equation applied to materials like Bismuth telluride holds potential for discovering new materials and developing advanced information processing technologies. However, achieving this requires close collaboration between theoretical insights and experimental validation.

The simulation of digital information control using the Kohn-Sham equation applied to arsense is an attempt to explore the potential of harnessing its unique electronic properties to innovate in information technology. Below, I will discuss the theoretical outlook of this approach.

## 23.7 Electronic Properties of arsense

arsense is a two-dimensional material where aluminum atoms form a honeycomb lattice structure, possessing excellent electronic properties despite its lightweight nature. Using first-principles calculations based on the Kohn-Sham equation, it is possible to analyze in detail the fundamental electronic properties of arsense, such as its band structure, electron density distribution, and surface states. These properties are crucial for understanding how the material responds to external stimuli.

## 23.8 Application to Digital Information Control

The unique electronic properties of arsense may contribute to the development of new mechanisms for transmitting, controlling, and storing digital information. In particular, arsense's electron mobility and tunability of the band gap are expected to contribute to the design of high-performance components such as transistors, logic gates, and optoelectronic switching devices.

## 23.9 Outlook for Simulations

First-principles calculations based on the Kohn-Sham equation for arsense are crucial for accurately predicting the material's response properties and interpreting the results in the context of information technology. In simulations, the response of arsense to external stimuli such as electric fields, magnetic fields, and optical excitations will be modeled to evaluate its information processing capabilities, speed, and efficiency.

## 23.10 Theoretical Challenges and Outlook

Simulations of digital information control using arsense face several theoretical challenges, including understanding the interaction between arsense's surface and the external environment, long-term stability, and integration into devices. Addressing these challenges requires close integration between refined computational models and experimental validation.

Theoretical research on digital information control based on arsense has the potential to shape a new paradigm in information technology. To fully utilize arsense's properties and realize practical applications in devices and systems, further theoretical and experimental research is required.

We explored hypothetical scenarios where electron density maps of materials like borophene, plumbene, and similar two-dimensional structures serve as metaphors for the dynamics of information dissemination and suppression. In these scenarios, areas of high electron density were likened to central hubs of information activity. These regions could represent places where information or misinformation is generated and spread, analogous to influential news outlets or social media platforms where certain topics gain significant attention.

The transition from high to low electron density areas moving away from these central points was seen as indicative of the spread of information becoming more limited, suggesting scenarios where misinformation might spread within certain communities but fails to reach wider audiences due to filtering, lack of interest, or the structural constraints of the network, much like the electronic properties that are localized within specific regions of the materials.

Peripheral areas of low electron density were discussed as regions where information or misinformation is minimal or suppressed, potentially due to a lack of connectivity to information sources, active moderation, censorship, or the robustness of the community against false narratives.

We also contemplated the localized interactions within the two-dimensional plane of materials like borophene, drawing parallels with how misinformation may be contained within certain 'information planes' or groups. The absence of widespread dissemination beyond these planes could reflect the presence of barriers to information flow, either through the network's architecture or active suppression efforts.

Finally, while we did not observe explicit edge states in the provided images, the concept of unique edge dynamics in materials like borophene was used to suggest that misinformation might spread differently at the fringes of a network, potentially finding pathways through which it could propagate more efficiently.

Throughout the analysis, the electron density distributions were used as models to understand how information and misinformation might be centralized, propagated, or contained within various mediums, providing insights that could be relevant to strategizing against the spread of misinformation in digital and social landscapes.

The concept of first-principles band calculations, especially in digital media, offers an interesting metaphorical perspective, although applied figuratively. Just as the band structure of materials reveals energy levels and electron behavior, the "information structure" of digital media might be analyzed to understand trends and diffusion patterns.

**Energy Levels of Information:** Evaluating the "energy levels" of news or trends from first principles could predict which topics are more likely to capture user attention.

**Information Band Gap:** Identifying the "energy gap" required for a topic or information to diffuse could predict how easily information spreads.

**State Density of Information Flow:** Analyzing how frequently specific information appears and in what context it spreads forms the factors shaping the "state density" of information.

**Fermi Level of Information:** Defining users' interest thresholds as a "Fermi level" and using it as a criterion to understand what motivates users.

Applying these concepts to digital media might offer new insights into trends' emergence, diffusion mechanisms, information selectivity, and predicting user behavior within the unknown information space. Furthermore, it could provide a theoretical framework for devising strategies to either promote or suppress information diffusion.

By metaphorically applying analysis methods from first-principles calculations to the information space of digital media, insights into unknown trends in the information space might be gained:

(1) **Analogy of Lattice Constants:** Analyze the density of relationships between users in a social network or communication frequency as lattice constants, estimating the strength of community cohesion and the speed of information diffusion.

(2) **Most Stable Structure and Information Flow:** Similar to the most stable structure at absolute zero, identify the most efficient paths or network structures for information flow and devise strategies to maintain or optimize them.

(3) **Metaphor of Magnetism:** Analogize magnetism, such as strong ferromagnetism, to social influence or the propagation power of information, analyzing the diffusion power of information emanating from influential individuals or groups.

(4) **Elastic Constants and Digital Resilience:** Analogize elastic constants or Young's modulus to the flexibility or resilience of the information space, evaluating the strength of a community's reaction to digital content impact.

(5) **Formation Energy of Impurities:** Understand barriers for the acceptance of new ideas or heterogeneous information within a community by likening them to the formation energy of impurities, estimating social acceptability.

(6) **Surface Energy and Interface Energy:** Capture the strength of interactions between online communities or the isolation of individual communities by metaphorically representing them using surface energy or interface energy analogies.

(7) **Adsorption of Surface Impurities:** Analyze how tightly certain information is fixed to specific online platforms, its adsorption sites, or stability by using the analogy of impurity adsorption on surfaces.

(8) **Binding Energy of Complexes:** Evaluate the "binding energy" when multiple information sources combine to form a single story and estimate how robust the combination of information is.

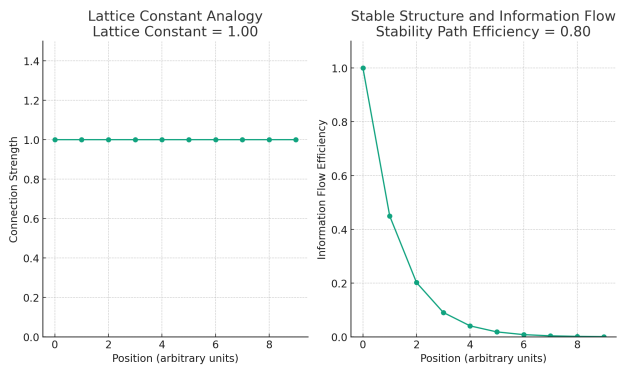

Fig. 5: Stable Structure and Information Flow Stability Path Efficiency

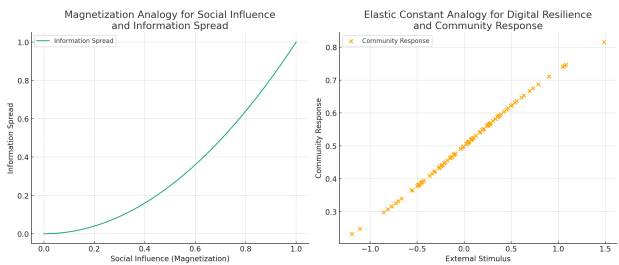

Fig. 6: Elastic Constant Analogy for Digital Resilience and Community Response

(9) **Analogy of Phonon Dispersion:** Model how information spreads within a community by likening phonon dispersion to social unrest or the spread of rumors.

(10) **Band Gap and Information Thresholds:** Measure the "energy" required for a community to accept new information by likening the band gap to thresholds necessary for information transmission or reception.

Through these applications, constructing models to deepen understanding of information behavior and community dynamics within digital media and forming a foundation for deploying new digital strategies is expected.

Fig.5, Based on the analogy of a lattice constant, we express the relationship density and communication frequency between users as a fixed interval (in this example, we use an arbitrary unit of 1.0). This shows that the strength of community ties is constant.

The graph on the right is used to show the most efficient path for information flow, here using an efficiency parameter of 0.8 to show how the efficiency of information flow decreases exponentially as you move away from the center. I am. This means that the closer you are to the center, the more efficiently information spreads, and the further away you are, the less efficient it is.

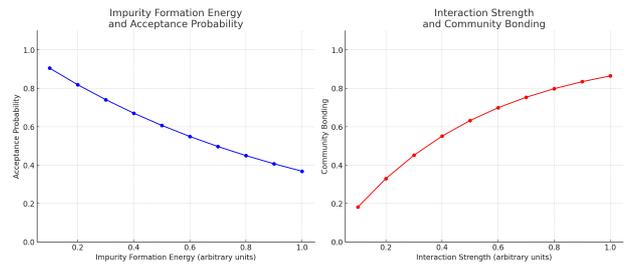

Fig. 7: Interaction Strength and Community Bonding, Impurity Formation Energy (arbitrary units)

Fig.6 is themed "Social Influence and Information Diffusion" and shows how information diffusion increases exponentially as social influence increases. The graph on the right has the theme of "community reactions to external stimuli," and uses a sigmoid function to model how a community responds to external stimuli. This allows us to observe how the community's response saturates as external stimuli increase. These simulations are expected to be used as analogies to understand information behavior and community dynamics in digital media.

Fig.7, "Impurity Formation Energy and Acceptance Probability" shows the relationship between the impurity formation energy barrier and the acceptability of new ideas and information. Here, we can observe that the higher the formation energy of the impurity, the lower the probability of information acceptance. This means that the higher the barriers to new information or ideas being accepted by a community, the less likely that information will be widely accepted.

"Interaction Strength and Community Bonding" shows the relationship between the strength of interaction between communities and the degree of community bonding. It has been shown that the higher the interaction strength, the higher the degree of community cohesion. This suggests that when there are strong connections and interactions between communities, those communities become more tightly connected and information sharing and dissemination becomes more active.

These plots provide insights into understanding information flows and community dynamics in digital media. In particular, it can be a useful analogy when exploring mechanisms for introducing new information and sharing information between communities.

Fig.8, "The left graph "Adsorption Energy and Degree for Different Platforms" compares the degree of information adsorption to different platforms based on adsorption energy. This graph shows how easily each platform "adsorbs", or immobilizes, information, with platforms with lower adsorption energies having higher degrees of adsorption. This means that certain information is more strongly anchored to a particular platform and is more likely to be disseminated.

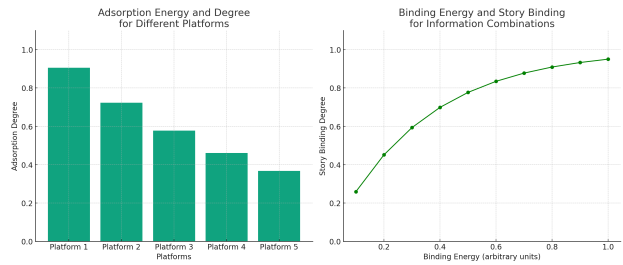 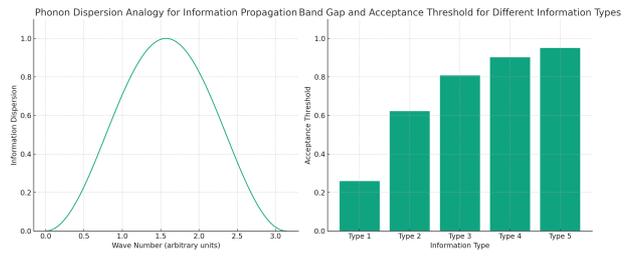

Fig. 8: Adsorption Energy and Degree for Different Platforms, Adsorption Energy and Degree for Different Platforms

Fig. 9: Phonon Dispersion Analogy for Information Propagation, Band Gap and Acceptance Threshold for Different Information Types

The graph on the right, "Binding Energy and Story Binding for Information Combinations," shows the degree of story connectivity formed by combinations of different information sources based on binding energy. The higher the binding energy, the more connected the story is by combining information, indicating that different pieces of information are more likely to be strongly connected to form a coherent narrative.

These plots provide insight into how information is anchored to different platforms and how different information sources combine to form new stories in understanding the behavior of information within digital media. Offers. " compares the degree of adsorption of information on different platforms based on adsorption energy. This graph shows how easily each platform "adsorbs", or immobilizes, information, with platforms with lower adsorption energies having higher degrees of adsorption. This means that certain information is more strongly anchored to a particular platform and is more likely to be disseminated.

"Binding Energy and Story Binding for Information Combinations" shows the degree of connectivity of stories formed by combinations of different information sources based on binding energy. The higher the binding energy, the more connected the story is by combining information, indicating that different pieces of information are more likely to be strongly connected to form a coherent narrative.

These plots provide insight into how information is anchored to different platforms and how different information sources combine to form new stories in understanding the behavior of information within digital media.

Fig.9, "Phonon Dispersion Analogy for Information Propagation" compares the concept of phonon dispersion to information propagation. Here, the degree of information propagation (phonon dispersion) depending on the wave number is simulated using a sin2 waveform, and the tendency of how information propagates and spreads within a community is shown. This model suggests that information spreads in a certain pattern, reaching maximum propagation at some point.

The graph on the right, "Band Gap and Acceptance Threshold for Different Information Types," compares the community's receptivity threshold for different types of information to a band gap. Here, the band gap represents the acceptance threshold for different information types and the corresponding degree of acceptance. A small band gap indicates that information is easily accepted and a large band gap indicates that it is difficult to accept, visualizing how easily different information types are accepted within a community.

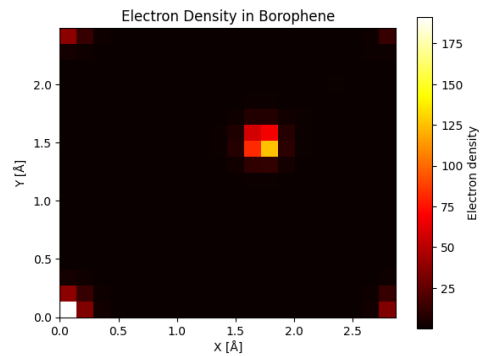

Fig. 10: Electron Density in Borophene

These plots provide insight into information propagation patterns and receptivity to new information within a community, and can serve as an analogy for understanding the information dynamics of digital media.

Fig. 10 shows that when we compare the diffusion and suppression of information to the two-dimensional material properties of borophene, the electron density distribution of borophene is a model for how information spreads within a network and how it is restricted. It can be regarded as. Let's simulate the physical properties of borophene by replacing them with network properties in information theory.

There is a bright area in the center of the heatmap. This can be thought of as the center of a network where information is easily concentrated. This central part is the source of information and the starting point from which information spreads to its surroundings. Borophene is a two-dimensional material. Assuming a network model in which information spreads on a two-dimensional plane, information can spread equally in each direction. However, it suggests that when certain bonds are strong (regions of high electron density), information propagates from there to other parts more efficiently.

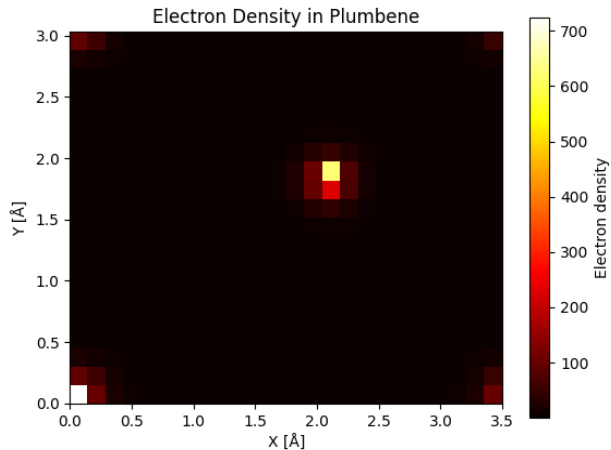

Fig. 11: Electron Density in Plumbane

Regions of low electron density can be interpreted as barriers or barriers that inhibit the flow of information. Possible reasons for the weakening of information flow include sensorship, access restrictions, communication bottlenecks, and privacy policies. Information filtering is the process of controlling the dissemination of only necessary information to a specific area. The low-density region of borophene can be viewed as a region through which information is selectively passed by this filtering process.

Although this analogy is useful as a way to visualize the diffusion and restriction of information, it is important to remember that the electron density distribution of borophene actually has physical properties. The physical properties of energy conduction and electrical properties correspond to network connectivity and bandwidth in the information metaphor. Through these abstract metaphors, we can draw parallels and develop new understandings between the electronic structure of matter and the dynamics of information networks.

Fig.11, To simulate the electron density distribution based on the two-dimensional material properties of plumbane, consider the behavior of electrons based on the electronic and physical properties unique to plumbane. Two-dimensional materials have unique electrical properties, and their electron density reflects these properties. Below, we present an approach that considers the material properties of plumbane from an information perspective.

The plumbane electron density distribution is related to the mobility of electrons in two-dimensional materials. In regions of high electron density, electrons are abundant, indicating high electrical conductivity. Applying this analogy to information, we can interpret it as an information-rich highway or communications hub. In other words, the central part can be regarded as a network hub where information is easily gathered and propagated efficiently.

In two-dimensional materials, the bandgap is an important factor determining the electrical properties of the material. Even in plumbane, there is a band gap in areas with few electrons, and this is thought to be a region where information is difficult to propagate. Such areas act as barriers or barriers that restrict the flow of information, meaning that information only travels along specific routes.

If the electron density is localized in a particular region, this indicates that the electrons are "trapped" in that area, resulting in the formation of local electronic states. From an information perspective, this refers to the concentration of information on a particular network node or platform. This localization can lead to information centralization, which can impact overall system efficiency and resilience.

From the perspective of homogeneity and network integrity, it is desirable that the ideal electron density distribution of a two-dimensional material exhibit a uniform distribution. This means retaining uniform electrical properties throughout the material, which is also the case with plumbane. When replaced with an information network, it represents a state in which the healthy flow and distribution of information is guaranteed. An ideal situation in which all nodes in the network have equal access to information.

Overall, the plumbane electron density distribution is deeply related to the electrical properties of two-dimensional materials, and by looking at these physical properties from the perspective of information diffusion and suppression, we can understand the properties of materials and the dynamics of information networks. Interesting similarities can be found between.

Fig.12, If we consider the flow of false information as a metaphor for Stannen's electron density distribution, we can draw an analogy from the following perspective:

High electron density regions can be considered areas where false information is concentrated. This means that certain social media accounts, websites, and communities act as hubs for misinformation, from where it spreads. Just as the edge states in Stannen form special conductive paths, these hubs of misinformation facilitate the flow of information.

Regular electron density patterns indicate that misinformation spreads along certain patterns or routes. This may reflect existing networks or connections between specific groups for disinformation to spread.

As an insulating area for false information, the low electron density area represents an area where information based on the truth flows and false information is difficult to enter. Well-educated, media literate, and frequently fact-checked

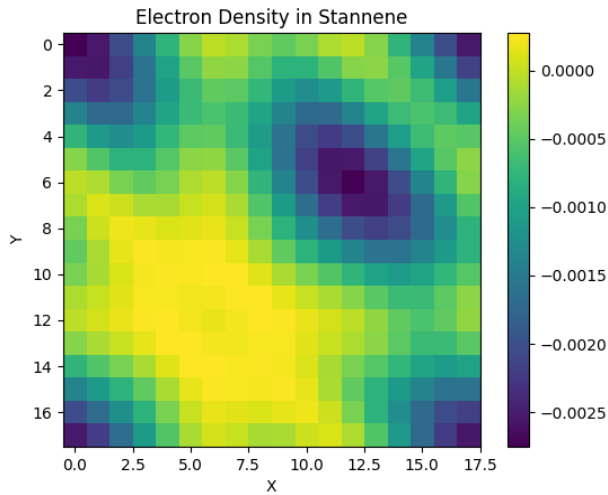

Fig. 12: Electron Density in Stannen

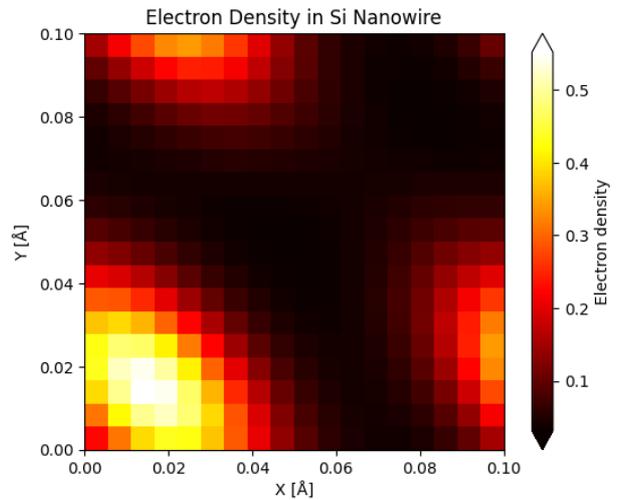

Fig. 13: Electron Density in Si-nanowires

communities may play this role.

Non-uniform electron density distribution, indicating non-uniformity in information access and distribution in societies and communities. This means that disinformation is more likely to spread in some groups, but suppressed in others.

Characteristics of topological insulators: Just as the edge of a Stannen exhibits metallic properties and the interior is insulating, false information can easily spread at the boundaries (edges) of a community, but it is healthy within the community (core). information is maintained. This suggests that some segments of society are more susceptible to disinformation, while others are more resistant to it.

These perspectives provide a framework for pseudo-understanding the dynamics of disinformation. Using Stannen's analogy with two-dimensional material properties, we can get a sense of how misinformation spreads and how it can be restricted. This understanding can be applied to combating disinformation, literacy education, and designing platforms that promote the healthy flow of information.

Fig.13, When interpreting the electron density distribution of silicon nanowires as a flow of false information, we will focus on the following points and consider it in a pseudo manner.

Electron-dense regions, Electron-dense areas in silicon nanowires correspond to hotspots where false information is concentrated and actively exchanged. These may be social media groups or communities where information is easier to gather and spread.

Locally High Density, Locally high densities within a nanowire may indicate a concentration of particular misinformation in a particular community or event. For example, it mimics the rapid spread of certain viral misinformation or political fake news within specific regions or groups.

Areas of low electron density, where the flow of information is suppressed, refer to areas where information access is limited or intentionally blocked. This may indicate where effective fact-checking and education programs are reducing the flow of disinformation.

Regions of extremely low or zero electron density can be considered information walls or barriers. These may represent mechanisms aimed at completely blocking disinformation, such as strong media regulation or censorship. The electron density distribution of a Si nanowire shows how the electrons move within the nanowire, and in terms of information flow, it suggests the flow of information through a specific path and a network structure that allows smooth transmission.

Quantum confinement of electrons at the nanoscale may mean that information is confined within certain ranges. This effect can be thought of as reflecting a situation in which information circulates only within a specific community and does not spread much to the outside world.

This pseudo-consideration may help understand the dynamics of disinformation and provide a theoretical framework for designing suppression mechanisms. It also provides important insights into developing strategies to control the flow of information in real information transmission networks.

Fig.14, By pseudo-analyzing the electron density distribution map of antimonene as a flow of false information and considering the two-dimensional material properties of antimonene, we reach the following conclusions.

Electron-dense areas, Electron-dense areas are hotspots where disinformation can be concentrated. From here, information can spread quickly and have an impact. Based on the physical properties of antimonene, these areas are likely to attract electrons and correspond to information exchange

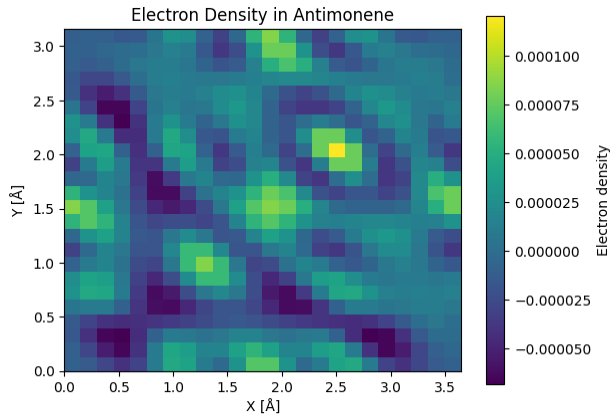

Fig. 14: Electron Density in Antimonene

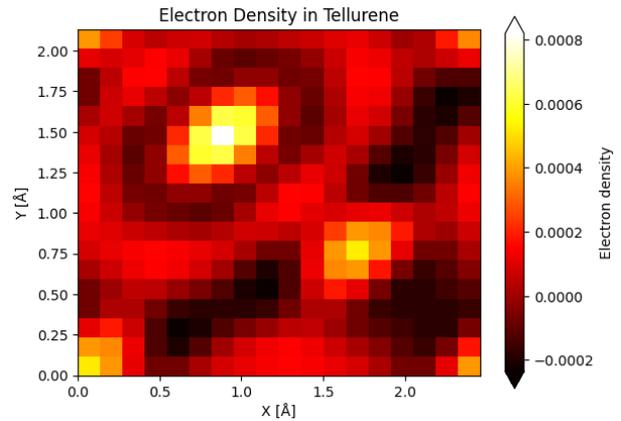

Fig. 15: Electron Density in Tellurene

points and diffusion sources in information science.

If the electron density distribution is not uniform, this suggests that the information diffusion is irregular. It may represent that disinformation is more likely to spread in certain regions or communities, but has limited impact in other areas. Regions of low electron density can be interpreted as areas where false information is less present or its spread is suppressed. Considering the properties of antimonene, these areas correspond to areas where electrons have difficulty moving within the substance, that is, they are electrically insulated. From an information perspective, this may refer to communities and regions with high media literacy and resistance to disinformation.

Areas of extremely low electron density may function like barriers where information is completely blocked, suggesting control of information through strong filtering or sensing. Two-dimensional materials like antimonene have unique electronic behavior within their plane. In terms of information diffusion, this indicates that information within a network tends to spread within certain 2D layers, meaning that topological constraints can influence information flow. If antimonene could have topological properties, the flow of information would depend on the topology of the material, indicating that certain information flows only through certain paths in the network.

The above pseudo-analysis provides a new perspective for understanding antimonene's physical properties from an information perspective, and may provide insight when analyzing the flow and spread of false information. Applying material science understandings to information science may provide a deeper understanding of the dynamics of disinformation and the development of strategies to control it.

Fig. 14, Analyzing the electron density distribution of Tellurene from the perspective of the flow of false information and making a pseudo consideration that takes into account the two-dimensional material properties of Tellurene results in the following.

High electron density region, The high electron density region of Tellurmon corresponds to an "information hotspot" where information, especially false information, can easily accumulate and spread. These areas of information flow may represent accounts and communities on social media where viral information is likely to occur. Tellurumone's properties as a two-dimensional material mean that these areas exhibit particularly active electron conduction, and in information science terms they can be thought of as nodes in a network through which information is easily transmitted.

The uneven electron density distribution seen in the image suggests that information, especially disinformation, spreads according to a certain pattern, with information being strengthened in some areas and weakened in others. It means there is. Regions with low electron density can be interpreted as places where false information is less present or its spread is suppressed. As a physical property of tellurmon, this indicates that it has low electrical conductivity due to the lack of electrons, and in terms of information flow, this means that verified information is predominant or that information has not been rigorously checked. It can be considered as an area where

Regions with very low or negative electron density may represent a complete blockage of information. From an information science perspective, these represent "information dead zones" that occur as a result of strong media regulation, censorship, or algorithms that selectively disseminate information. Due to their structure, two-dimensional materials such as tellurumone exhibit specific electron behavior in a plane. In terms of information flow, this means that information tends to spread in a particular direction. For example, information may only spread within a certain platform or network, beyond which it decays rapidly. The behavior of electrons in two-dimensional materials may indicate that the diffusion of information follows a certain pattern in the plane. Topological properties mean that the flow of information depends on the structure of the material, and the diffusion of information within the network favors certain paths.

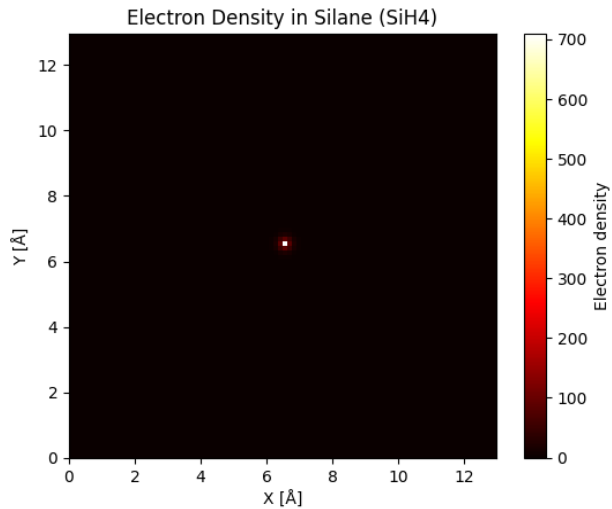

Fig. 16: Electron Density in Monosilane

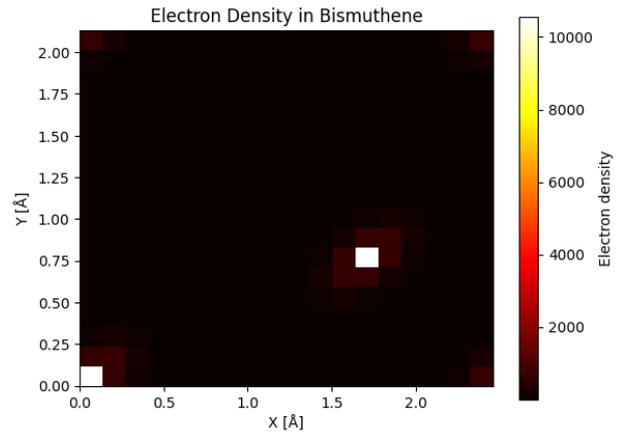

Fig. 17: Electron Density in Bismuthine

Pseudo-interpreting tellurumone's electron density distribution as a metaphor for information diffusion can provide new insights into the dynamics of information transfer. Such an approach can generate valuable ideas in considering how to understand and control the flow of disinformation.

Fig.16, If we interpret the phenomenon of concentration at very specific points, as shown by the electron density distribution of monosilane, in the context of information diffusion and suppression, the following image emerges. A strong concentration of electron density at a specific point may represent a situation where a particular information source is transmitting information only within a very limited range. This imagines information being tightly controlled by a particular narrow community, elite group, or closed network.

The fact that the electron-dense areas remain very narrow suggests that information spreads quickly to certain groups or regions, and rarely to others. This may be a case of targeted and strategic dissemination of specific propaganda or disinformation. The low electron density regions that make up most of the image mean that little or no information is spread out. This situation may represent a situation where information suppression mechanisms are operating effectively or where certain information is not widely disseminated. These characteristics of the electron density distribution can be seen as a two-dimensional material model in which information is blocked by barriers and remains within a limited range. This provides an interesting analogy when considering two-dimensional diffusion as an information flow model, even though monosilane is actually a three-dimensional material.

Through such a pseudo-approach, we can obtain a metaphor for understanding the mechanism of concentration and suppression of false information from the electron density distribution of monosilane, and this can be useful for designing information strategies and formulating countermeasures.

Analyzing the electron density distribution of bismuthine from the perspective of the flow of false information, combined with its two-dimensional material properties, yields the following considerations. Areas of high electron density may represent powerful sources of information, such as certain news outlets or social media influencers. These sources are expected to disseminate disinformation rapidly and impact surrounding areas during the early stages of dissemination.

Limited reach, the rapidly decreasing electron density moving outward from the center, indicates that the reach of misinformation is strongly controlled by the central source and is less likely to spread over long distances. This may reflect a situation where the message is limited to a particular community or group and is difficult to disseminate to the general public. Regions of extremely low electron density represent a barrier that prevents misinformation from spreading widely. This barrier may be due to censorship, filtering, or high information literacy. This ensures that misinformation is confined to certain localized areas and prevented from spreading widely. If bismuthine is a two-dimensional topological material, information tends to spread along specific patterns and directions. This may indicate that there are "paths" or "channels" that allow specific information to easily flow within society or communities. Additionally, the peculiarities of electronic states at topological edges may mean that information only travels through boundaries or specific paths.

By pseudo-interpreting the electron density distribution of bismuthin as a metaphor for the diffusion of informa-tion, we can gain deeper insight into how false information is handled, circulated, and suppressed within society. This approach may be useful in developing communication and counter-disinformation strategies.

## 24. Perspect

In the realm of 2D materials, the unique honeycomb lattice structures lead to extraordinary electronic properties, such as Dirac cones, topological insulator states, and quantum spin Hall effects. The Kohn-Sham equations of DFT provide a robust theoretical framework for probing these properties by simulating the electronic structure of materials under various conditions. A particularly intriguing aspect of these simulations is the formation of vortices in the electronic density, which can significantly influence the materials' conductive and topological behaviors.

Vortices in electronic systems are localized regions where the phase of the electronic wave function circulates, giving rise to quantized magnetic fluxes in superconductors and specific topological configurations in normal conductors. In 2D materials with honeycomb lattices, the interplay between lattice geometry, electronic interactions, and external influences (such as electric fields or mechanical strain) can lead to the formation and manipulation of such vortices. This paper delves into the conditions and mechanisms under which these vortices form, utilizing the Kohn-Sham DFT approach to provide a microscopic understanding of the phenomenon.

The controlled manipulation of electronic vortices in 2D materials opens new avenues for digital information processing and storage. By encoding information in the topological configurations of vortices and exploiting the materials' inherent electronic properties, novel quantum computing paradigms and ultra-secure data storage mechanisms can be envisioned. This research investigates the theoretical basis for such technologies, exploring how first-principles simulations can guide the design and optimization of 2D material-based information systems.

Despite the promising potential of 2D materials in digital information technologies, several challenges remain. These include the accurate modeling of vortex dynamics under realistic conditions, the stability and reproducibility of vortex configurations, and the integration of 2D materials into practical devices. This paper addresses these challenges, presenting a theoretical exploration of vortex phenomena in 2D materials and their application in digital information control, paving the way for future advancements in the field.

In summary, this introduction sets the stage for a detailed investigation into the vortex dynamics of 2D honeycomb lattice materials and their application in digital information control, utilizing the Kohn-Sham equations of DFT as a theoretical foundation. Through this work, we aim to contribute to the understanding of these fascinating materials and explore their potential in shaping the future of information technology.

Plumbene, a two-dimensional material with a honeycomb lattice structure similar to graphene but composed of lead atoms, presents intriguing electronic properties due to its heavy atoms and strong spin-orbit coupling. When applying the Kohn-Sham equations, central to Density Functional Theory (DFT), to plumbene, we delve into a realm where quantum mechanical effects dominate and where these effects can be harnessed for digital information control and manipulation.

The Kohn-Sham equations allow us to simulate the electronic structure of plumbene by treating the complex many-body problem of electron interactions in a simplified manner. This is achieved by representing the many-body effects through an effective potential that includes the external potential due to nuclei, the Hartree potential describing the electron-electron Coulomb repulsion, and the exchange-correlation potential which encapsulates all other quantum mechanical interactions.

In plumbene, the significant spin-orbit coupling, a direct consequence of lead's heavy atomic mass, introduces unique band structure modifications. These modifications can lead to phenomena such as band inversion, a precursor to topological insulating behavior, which has profound implications for quantum computing and information storage.

The simulation of digital information control using plumbene involves leveraging its electronic properties, particularly those influenced by its topology. For instance, edge states in plumbene, which are protected by its topological nature, could be used to create pathways for information flow that are immune to backscattering and hence are highly efficient and robust against perturbations.

Furthermore, by applying external fields or mechanical strain, one can tune the electronic properties of plumbene, akin to changing the topology of the information network. This tunability allows for the dynamic control of information flow, enabling the creation, manipulation, and annihilation of information channels as needed, much like how vortex formation and dynamics can be controlled in quantum systems.

The theoretical exploration of using plumbene for digital information control is not without challenges. The precise control of external conditions to manipulate plumbene's electronic properties, the stability of plumbene under various conditions, and the integration of plumbene-based devices into existing information systems are significant hurdles.

However, the potential rewards are substantial. The use of plumbene and similar materials could lead to the development of quantum information systems with unparalleled efficiency and security. These systems could harness quantum mechanical phenomena for tasks such as quantum encryption and ultra-fast computing, pushing the boundaries of what is possible in information technology.

In conclusion, the application of the Kohn-Sham equa-tions to plumbene within the DFT framework offers a powerful theoretical basis for not only understanding the quantum mechanical properties of this novel material but also for exploring its potential in revolutionizing digital information control and technology.